\documentclass[aps,pre,reprint,superscriptaddress]{revtex4-1}

\pdfoutput=1

\usepackage{graphicx}
\usepackage{amssymb}
\usepackage{xcolor}
\usepackage{pdfpages}
\usepackage{pgffor}
\usepackage{CJK}

\makeatletter
\AtBeginDocument{\let\LS@rot\@undefined}
\makeatother

\begin{document}

\begin{CJK*}{UTF8}{} % Use default fonts from CJK (see below)

\title{Stratification in Drying Films Containing Bidisperse Mixtures of Nanoparticles}

\author{Yanfei Tang ({\CJKfamily{gbsn}唐雁飞})}
\affiliation{Department of Physics, Center for Soft Matter and Biological Physics,
and Macromolecules Innovation Institute, Virginia Polytechnic Institute and State University,
Blacksburg, Virginia 24061, USA}
\author{Gary S. Grest}
\affiliation{Sandia National Laboratories, Albuquerque, NM 87185, USA}
\author{Shengfeng Cheng ({\CJKfamily{gbsn}程胜峰})}
\affiliation{Department of Physics, Center for Soft Matter and Biological Physics,
and Macromolecules Innovation Institute, Virginia Polytechnic Institute and State University,
Blacksburg, Virginia 24061, USA}
\email{chengsf@vt.edu}

\date{\today}

\begin{abstract}
Large scale molecular dynamics simulations for bidisperse nanoparticle suspensions with an explicit solvent are used to investigate the effects of evaporation rates and volume fractions on the nanoparticle distribution during drying. Our results show that ``small-on-top'' stratification can occur when ${\rm Pe}_s\phi_s \gtrsim c$ with $c\sim 1$, where ${\rm Pe}_s$ is the P\'{e}clet number and $\phi_s$ is the volume fraction of the smaller particles. This threshold of ${\rm Pe}_s\phi_s$ for ``small-on-top'' is larger by a factor of $\sim\alpha^2$ than the prediction of the model treating solvent as an implicit viscous background, where $\alpha$ is the size ratio between the large and small particles. Our simulations further show that when the evaporation rate of the solvent is reduced, the ``small-on-top'' stratification can be enhanced, which is not predicted by existing theories. This unexpected behavior is explained with thermophoresis associated with a positive gradient of solvent density caused by evaporative cooling at the liquid-vapor interface. For ultrafast evaporation the gradient is large and drives the nanoparticles toward the liquid/vapor interface. This phoretic effect is stronger for larger nanoparticles and consequently the ``small-on-top'' stratification becomes more distinct when the evaporation rate is slower (but not too slow such that a uniform distribution of nanoparticles in the drying film is produced), as thermophoresis that favors larger particles on the top is mitigated. A similar effect can lead to ``large-on-top'' stratification for ${\rm Pe}_s\phi_s$ above the threshold when ${\rm Pe}_s$ is large but $\phi_s$ is small. Our results reveal the importance of including the solvent explicitly when modeling evaporation-induced particle separation and organization and point to the important role of density gradients brought about by ultrafast evaporation.
\end{abstract}

\maketitle

\end{CJK*}

\noindent Evaporation is a ubiquitous process that plays an important role in many diverse fields including climate, environment, and industry.\cite{Persad2016} It is also frequently used in material fabrications. For example, controlled evaporation is used to make polymer thin films,\cite{ChenNL2007,Strawhecker2001} polymeric particles,\cite{ShinACSNano2017} and nanocomposites,\cite{Jouault2014,ArciniegasNL2016} and to assemble building blocks including particles into superstructures.\cite{Bigioni2006,DickeyACSNano2008,ChenACSNano2009,Utgenannt2016,Reinhart2017,Cheng2017} In a relatively simple case where a suspension containing particles undergoes drying, the structure of the final dry film is determined by two competing factors: the diffusion of the particles and the receding motion of the liquid/vapor interface induced by solvent evaporation.\cite{Routh2004,Routh2013} The competition is quantified by a dimensionless P\'{e}clet number, ${\rm Pe} = H v/D$, where $H$ is the thickness of the interfacial region affected by evaporation and can be taken as the film thickness for thin films, $D$ is the diffusion constant of the particles, and $v$ is the receding speed of the interface. Routh and Zimmerman derived the governing equation for the evolution of particle volume fractions, which is referred to as the RZ model hereafter, and obtained numerical solutions at various P\'{e}clet numbers.\cite{Routh2004} Their analyses showed that when ${\rm Pe} \gg 1$, the particles are trapped and accumulated near the liquid/vapor interface, forming a skin layer since their diffusion is slow compared to the recession of the interface. However, when ${\rm Pe} \ll 1$, the diffusion of the particles is faster than the motion of the interface and the particles remain almost uniformly distributed in the drying film. 

Since diffusion constant of a particle depends on its size, the situation becomes particularly intriguing when the suspended particles are polydisperse. From the Einstein-Stokes relationship, $D=k_{\rm B}T/(3\pi \eta d)$, where $k_{\rm B}$ is the Boltzmann constant, $T$ the temperature, $\eta$ the solvent viscosity, and $d$ the particle diameter. The P\'{e}clet number is thus proportional to $d$. The simplest polydisperse system is a suspension containing particles of two sizes $d_l$ and $d_s$ with a size ratio $\alpha = d_l/d_s>1$. Trueman \textit{et al.} extended the RZ model to such bidisperse suspensions and combined numerical simulations and experiments to show that the larger particles accumulate while the smaller ones are depleted near the interface when ${\rm Pe}_l > 1 > {\rm Pe}_s$.\cite{Trueman2012, Trueman2012a} This is called ``large-on-top'' stratification. Using the extended RZ model, Atmuri \textit{et al.} studied the effects of inter-particle interactions amongst the same species (i.e., the particles of the same size) on the particle distribution during drying of the suspension.\cite{Atmuri2012}. They also investigated suspensions containing particles of the same size but some of them are neutral while the others are charged. Their numerical simulations showed that in this case the charged particles are depleted at the receding interface because of the repulsion between likely charged particles. As a result, neutral particles remain and accumulate near the interface. This finding is consistent with an earlier study of Nikiforow \textit{et al.},\cite{Nikiforow2010} who studied a latex blend of charged and neutral particles of roughly the same size and found that stratification between the two species readily occurred after drying of the film with the neutral particles accumulating immediately below the film/air interface.

From these previous studies, it was believed that in bidisperse particle suspensions, stratification would likely be produced if the P\'{e}clet numbers of the two components were on different sides of unity (e.g., ${\rm Pe}_l > 1 > {\rm Pe}_s$) and the particles with a smaller diffusivity (e.g., the larger particles) would accumulate at the top of the dried film.\cite{Trueman2012, Trueman2012a} However, Fortini \textit{et al.} recently discovered the occurrence of a novel ``small-on-top'' stratifying scenario when ${\rm Pe}_l \gg {\rm Pe}_s \gg 1$.\cite{Fortini2016,Fortini2017} Namely the smaller particles accumulate near the interface when the evaporation is very fast for both large and small particles. They proposed that for very fast evaporation both large and small particles first accumulate just below the receding interface, creating gradients of their concentration distributions in the direction perpendicular to the film. The concentration gradients lead to gradients of the associated osmotic pressure, which cause the particles to drift. However, the drift velocity is asymmetric for the large and small particles. Fortini \textit{et al.} argued that if the volume fraction of small particles is large enough to make them the majority phase just below the film/air interface, then the large particles will drift away from this region faster than the small particles roughly by a factor of $\alpha^2-1$. The net result is an accumulation of small particles at the top of the drying film. The finding of Fortini \textit{et al.} seems to be consistent with a phenomenon observed earlier by Luo \textit{et al.}, who studied drying aqueous dispersions containing a mixture of latex particles with a diameter $\sim 550$ nm and much smaller ceramic nanoparticles (NPs) and found an enrichment of NPs in interstitial spaces among latex particles near the top surface of the drying film.\cite{Luo2008} Howard \textit{et al.} performed numerical simulations based on an implicit solvent model, similar to the one used by Fortini \textit{et al.}, to systematically study the effects of particle size ratios and evaporation rates on stratification.\cite{Howard2017} They found that ``small-on-top'' stratification can persist even when the P\'{e}clet numbers are of order 1 and noticed an unexpected accumulation of the larger particles near the substrate at small evaporation rates (i.e., small $v$).

To better understand stratifying phenomena, Zhou, Jiang, and Doi proposed a diffusion model, referred to as the ZJD model hereafter, for mixtures of hard spheres up to second virial coefficients.\cite{Zhou2017} The equations describing the time evolution of particle concentrations in the ZJD model are similar to those in the extended RZ model but the expressions for chemical potential are different.\cite{Trueman2012,Zhou2017} Analyses and numerical solutions of the ZJD model revealed that the ``small-on-top'' structure is created by the cross-interactions between particles of different sizes, which affect the larger particles much more strongly than the smaller ones roughly by a factor of $\alpha^3$.\cite{Zhou2017} A state diagram in the ${\rm Pe}_s$--$\phi_s$ plane was predicted where stratification occurs if $\alpha^2(1+{\rm Pe}_s)\phi_s > 1$, with $\phi_s$ being the initial volume fraction of the smaller particles.

Makepeace \textit{et al.} recently combined experiments and simulations to test the ZJD model.\cite{Makepeace2017} They found that at low particle concentrations the ZJD model fit their measurements and modeling data reasonably well while for concentrated suspensions the ZJD model significantly over-predicts ``small-on-top'' stratification, i.e., actual stratification occurs at $\alpha$, ${\rm Pe}_s$, and $\phi_s$ much larger than those predicted by the ZJD model. Liu \textit{et al.} performed experiments on the drying of suspensions containing a mixture of larger polystyrene NPs and smaller silica NPs and identified ``small-on-top'' states via atomic force microscopy (AFM) characterization of the film surface.\cite{Liu2018} Their results seem to fit the ZJD model, though their measurements are in the ${\rm Pe}_l > 1 > {\rm Pe}_s$ regime. Mart\'{i}n-Fabiani \textit{et al.} showed that stratification can be turned on and off on demand by mixing smaller particles, whose size can be varied by changing the pH of the suspension, and larger particles with a fixed size.\cite{Martin-Fabiani2016} At low pH, $\alpha \approx 7$ and ``small-on-top'' stratification occurs while at high pH, $\alpha$ decreases to about 4 and stratification is suppressed. Their work demonstrated the effects of high particle concentrations and the associated jamming that prevents the particles to stratify.

\begin{table*}[t]
\centering
\caption{Parameters for the fifteen systems studied.}
\begin{tabular}{cccccccccc}
\hline
System & $N_l$  & $N_s$   & $\phi_l$  & $\phi_s$  & $H/\sigma$     & $\zeta$ & $v\tau/\sigma$ & $\textrm{Pe}_{l}$ & $\textrm{Pe}_{s}$   \\ \hline
$\phi_{0.011}R_{30}$ & 200 & 1920 & 0.072 & 0.011 & 289.2 & 30 & 1.14$\times 10^{-3}$ & 109.7 & 27.4 \\
$\phi_{0.011}R_{5}$  & 200 & 1920 & 0.072 & 0.011 & 289.2 & 5 & 2.05$\times 10^{-4}$ & 19.7 & 4.9 \\
\hline
$\phi_{0.034}R_{30}$ & 200 & 6400 & 0.068 & 0.034 & 304.4 & 30 & 1.13$\times 10^{-3}$ & 114.3 & 28.6 \\
$\phi_{0.034}R_{5}$  & 200 & 6400 & 0.068 & 0.034 & 304.4 & 5 & 2.04$\times 10^{-4}$ & 20.7 & 5.2 \\
$\phi_{0.034}R_{1}$  & 200 & 6400 & 0.068 & 0.034 & 304.4 & 1 & 4.33$\times 10^{-5}$ & 4.4 & 1.1 \\
\hline
$\phi_{0.068}R_{30}$ & 200 & 12800 & 0.068 & 0.068 & 306.5 & 30 & 1.04$\times 10^{-3}$ & 105.9 & 26.5 \\
$\phi_{0.068}R_{5}$  & 200 & 12800 & 0.068 & 0.068 & 306.5 & 5 & 2.03$\times 10^{-4}$ & 20.8 & 5.2 \\
$\phi_{0.068}R_{1}$  & 200 & 12800 & 0.068 & 0.068 & 306.5 & 1 & 4.21$\times 10^{-5}$ & 4.3 & 1.1 \\
\hline
$\phi_{0.10}R_{30}$ & 200 & 19200 & 0.067 & 0.10 & 309.7 & 30 & 8.69$\times 10^{-4}$ & 89.7 & 22.4 \\
$\phi_{0.10}R_{5}$  & 200 & 19200 & 0.067 & 0.10 & 309.7 & 5 & 1.96$\times 10^{-4}$ & 20.2 & 5.1 \\
$\phi_{0.10}R_{1}$  & 200 & 19200 & 0.067 & 0.10 & 309.7 & 1 & 4.15$\times 10^{-5}$ & 4.3 & 1.1 \\
\hline
$\phi_{0.13}R_{30}$ & 200 & 25600 & 0.067 & 0.13 & 307.2 & 30 & 7.61$\times 10^{-4}$ & 78.0 & 19.5 \\
$\phi_{0.13}R_{5}$  & 200 & 25600 & 0.067 & 0.13 & 307.2 & 5 & 1.91$\times 10^{-4}$ & 19.6 & 4.9 \\
\hline
$\phi_{0.16}R_{20}$ & 200 & 32000 & 0.065 & 0.16 & 317.8 & 20 & 6.37$\times 10^{-4}$ & 67.5 & 16.9 \\
$\phi_{0.16}R_{5}$  & 200 & 32000 & 0.065 & 0.16 & 317.8 & 5 & 1.85$\times 10^{-4}$ & 19.6 & 4.9 \\
 \hline
\end{tabular}
\label{Table:system}
\end{table*}

From the reported studies we now understand that stratifying phenomena in a drying suspension containing a bidisperse mixture of neutral particles depend on several factors including the evaporation rate of the solvent, the initial volume fractions of the particles, the particle size ratio, and the interactions between the particles.\cite{Atmuri2012, Fortini2016, Zhou2017, Howard2017, Makepeace2017} The ZJD model predicts that for the initial volume fractions only that of the smaller particles matters.\cite{Zhou2017} However, in previous simulations\cite{Trueman2012,Trueman2012a,Atmuri2012,Fortini2016,Howard2017,Makepeace2017} and theory\cite{Zhou2017} the solvent was treated as an implicit, uniform viscous background. Very recently, Sear and Warren used the Asakura-Oosawa model to study the drift of a large particle in a solute (i.e., small particle) gradient,\cite{Asakura1954,Sear2017} taking into account the contribution from the solvent back-flow to the pressure gradient, and showed that the analyses of Fortini \textit{et al.} and the ZJD model based on an implicit solvent overestimate the drift velocity of large particles roughly by a factor of $\alpha^2$.\cite{Sear2017} With this correction, their prediction is that ``small-on-top'' stratification occurs only when ${\rm Pe}_s\phi_s \gtrsim 1$. Therefore, the threshold of ${\rm Pe}_s$ driving a system into the ``small-on-top'' regime at a given $\phi_s$ is higher than the ZJD prediction roughly by a factor of $\alpha^2$, which may provide an explanation of the finding of Makepeace \textit{et al.} that the ZJD model tends to over-predict stratification.\cite{Makepeace2017} Hereafter we refer the work of Sear and Warren as the SW model. In another recent work,\cite{Sear2018} Sear applied a gelation model originally developed by Okuzono and Doi for the drying of a polymer film \cite{Okuzono2006} to stratifying phenomena and considered the jamming of particles at high volume fractions and the resulting dynamic arrest of particle motion, which occur when the accumulation of particles near a receding liquid/vapor interface surpasses the jamming point. In the Sear model, ``small-on-top'' stratification only occurs for a finite range of $\phi_s$: $0.64/{\rm Pe}_s < \phi_s < 0.2$. These studies thus point to the importance of including a solvent explicitly when studying the drying of a particle suspension. Furthermore, all the theoretical models developed so far are based on isothermal systems but our previous work revealed that temperature and density gradients can emerge in a fast evaporating liquid,\cite{Cheng2011} whose roles in stratifying phenomena are unclear.

To fill the gap, here we report large-scale molecular dynamics (MD) simulations of the drying of bidisperse particle suspensions with an explicit solvent. To span regimes from ${\rm Pe}_l \gg {\rm Pe}_s \gg 1$ to ${\rm Pe}_l \gg 1 \gtrsim {\rm Pe}_s$, we use NPs with diameters 20 and 5 times of that of the solvent. In particular, we focus on the role of the evaporation rate (i.e., ${\rm Pe}_s$) and $\phi_s$ in controlling the distribution of NPs in the resulting dry films.

\bigskip
\noindent{\bf RESULTS AND DISCUSSION}

\noindent All of our simulations have $N_l=200$ large NPs (LNPs) of a diameter $d_l =20 \sigma$, where $\sigma$ is the unit of length. The small NPs (SNPs) have a diameter $d_s=5\sigma$ and their number, $N_s$, is varied from 1920 to 32000. Further details of the simulations are given in the Methods section. All the NPs are initially dispersed in a liquid consisting of Lennard-Jones (LJ) particles of a size $\sigma$ in equilibrium with its vapor in a rectangular box with dimensions $L_x \times L_y \times L_z$, where $L_x = 201\sigma$, $L_y = 201\sigma$, and $L_z = 477\sigma$. The number of LJ particles vary from about $7\times 10^6$ for $N_s=1920$ to about $5\times 10^6$ for $N_s=32000$. The initial volume fraction of NPs is $\phi_i \equiv \pi N_i d_{i}^{3}/(6 L_x L_y H)$, where $i \in \{l, s\}$ and $H$ is the film thickness at equilibrium. Evaporation of the solvent is implemented by removing the LJ particles in the deletion zone $[L_z -100\sigma, L_z]$.\cite{Cheng2011} The evaporation rate is controlled by setting the number of particles, $\zeta$, removed every $\tau$, where $\tau$ is the LJ unit of time. When the solvent evaporates into a vacuum, the evaporation rate is initially very high ($\zeta \sim 600$), then decreases with time, and finally reaches a plateau value corresponding to $\zeta \sim 20$--$30$, which slightly depends on $\phi_s$.\cite{Cheng2011} For the drying NP suspensions studied in this paper, we set $\zeta$ constant and the values of $\zeta$ are varied to span very fast evaporation ($\zeta =20$ or $30$) to the slowest evaporation rate ($\zeta = 1$) accessible with current computational resources.\cite{ACSNano2018Note1} We label each system as $\phi_{\phi_s}R_{\zeta}$ using the values of $\phi_s$ and $\zeta$. For each system we directly follow the location of the liquid/vapor interface during evaporation and compute $v$, which quantifies the speed of evaporation. For diffusion constants, we take $D_l = 3 \times 10^{-3} \sigma^2/\tau$ from our previous study \cite{Cheng2013} and assume $D_s = D_l d_l/d_s = \alpha D_l$. The values of P\'{e}clet numbers can then be estimated. In particular, ${\rm Pe}_s = \frac{Hv}{D_s} \approx \frac{v}{4\times 10^{-5}\sigma/\tau}$. For the systems studied here, $v \approx 4\zeta \times 10^{-5}\sigma/\tau$. As a result, the values of $\zeta$ and ${\rm Pe}_s$ are close, the latter of which can thus be roughly read from the subscript of $R$ in the system label $\phi_{\phi_s}R_{\zeta}$. Then ${\rm Pe}_l = \alpha {\rm Pe}_s = 4{\rm Pe}_s$. All the systems and parameters are listed in Table~\ref{Table:system}.

\begin{figure}[tp]
\includegraphics[width = 0.45\textwidth]{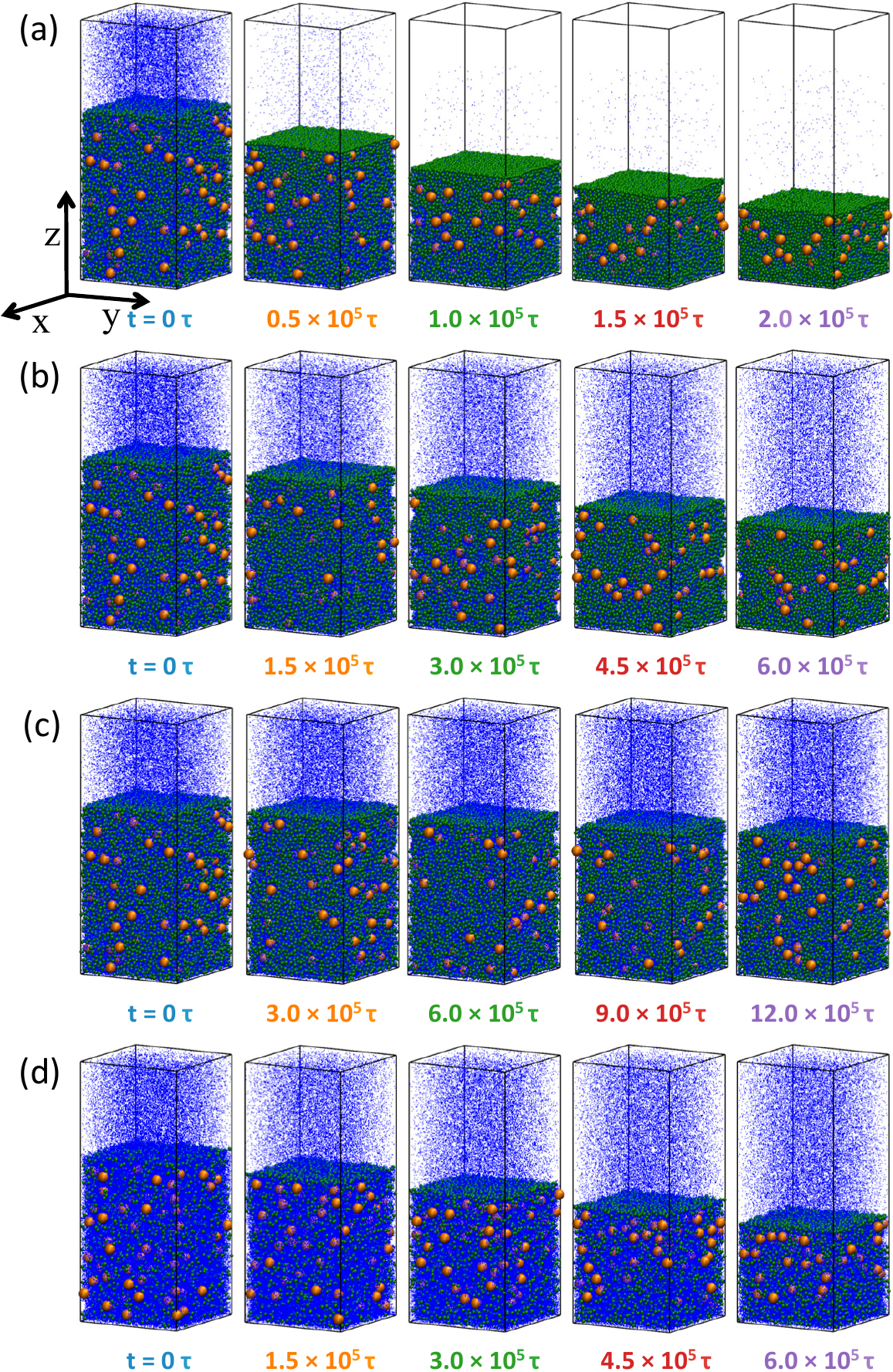}
\caption{Snapshots of the systems during evaporation for (a) $\phi_{0.10} R_{30}$, (b) $\phi_{0.10} R_{5}$, (c) $\phi_{0.10} R_{1}$, and (d) $\phi_{0.034} R_{5}$. Time is indicated below each snapshot with $t=0\tau$ for the equilibrium state prior to evaporation. Corresponding density profiles are plotted in Fig.~\ref{Figure:density}. Color code: LNPs (orange), SNPs (green), and solvent (blue). For clarity, only 5\% of the solvent beads are visualized. In the last frame the volume fractions of NPs are: (a) $\phi_l = 0.15$, $\phi_s = 0.23$; (b) $\phi_l = 0.11$, $\phi_s = 0.16$; (c) $\phi_l = 0.080$, $\phi_s = 0.12$; (d) $\phi_l = 0.11$, $\phi_s = 0.057$.}
\label{Figure:snapshots}
\end{figure}

Snapshots of 4 NP suspensions under various evaporation rates are shown in Fig.~\ref{Figure:snapshots}. These 4 are picked as they are representative to demonstrate how the distribution of NPs in a drying film changes when the evaporation rate (i.e., ${\rm Pe}_s$) and $\phi_s$ are varied. Snapshots of the other 11 systems can be found in the Supporting Information. We first focus on the systems $\phi_{0.10} R_{30}$, $\phi_{0.10} R_{5}$, and $\phi_{0.10} R_{1}$ with $N_s=19200$ [Figs.~\ref{Figure:snapshots}(a)-(c)] and investigate the role of evaporation rates. For the ultrafast evaporating system $\phi_{0.10} R_{30}$, the SNPs quickly accumulate and form a dense skin layer near the liquid/vapor interface, as shown in Fig.~\ref{Figure:snapshots}(a). However, the LNPs are also found to accumulate just below this interfacial layer of SNPs. When the evaporation rate is reduced by a factor of 6 in the system $\phi_{0.10} R_{5}$, the skin layer of SNPs becomes less distinct, though the SNPs still accumulate near the interface, as shown in Fig.~\ref{Figure:snapshots}(b). In this case the extent of accumulation for the LNPs just below the surface layer of SNPs diminishes significantly. As a result, the ``small-on-top'' stratification is enhanced in $\phi_{0.10} R_{5}$, in which the solvent evaporates more slowly than in $\phi_{0.10} R_{30}$. This trend is not predicted by the existing theories,\cite{Fortini2016, Zhou2017} which anticipate that ``small-on-top'' stratification should be suppressed and become less distinct when the evaporation rate is reduced. Below we will show that this unexpected behavior results from the density gradients of the solvent that develop during evaporation. When the evaporation rate is further reduced by a factor of 5 in the system $\phi_{0.10} R_{1}$, the accumulation of NPs near the interface almost disappears and the LNPs and SNPs are uniformly distributed in the drying film, as shown in Fig.~\ref{Figure:snapshots}(c).

Figure~\ref{Figure:snapshots}(d) shows the system $\phi_{0.034} R_{5}$, which has the same evaporation rate as $\phi_{0.10} R_{5}$ [Fig.~\ref{Figure:snapshots}(b)], but $\phi_s$ is smaller by a factor of 3. Compared to $\phi_{0.10} R_{5}$, the surface accumulation of SNPs during solvent evaporation is weaker and the enrichment of LNPs near the receding interface is much stronger in $\phi_{0.034} R_{5}$.

\begin{figure*}[t]
\includegraphics[width = \textwidth]{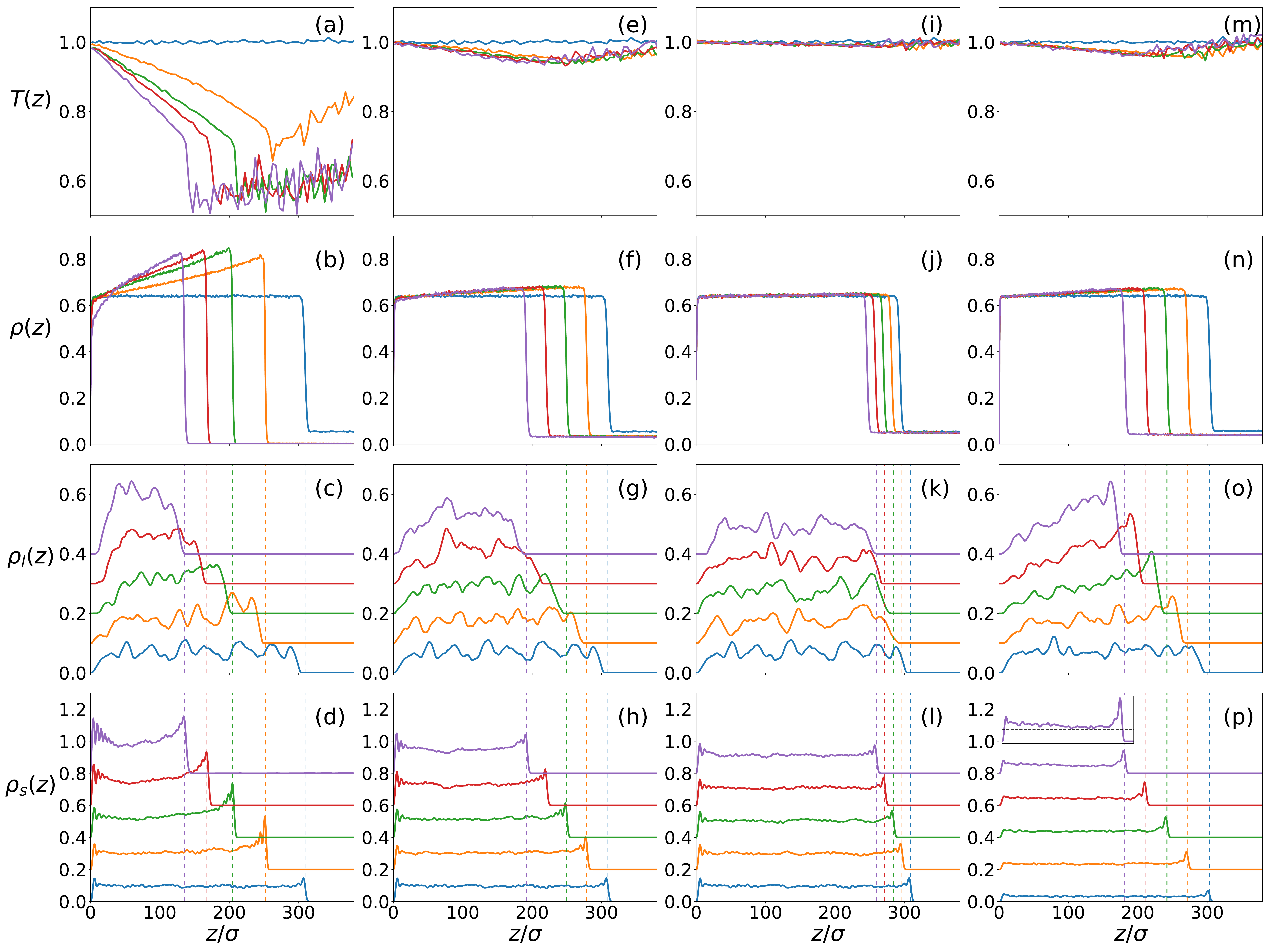}
\caption{Temperature profile during evaporation (top row) and density profiles for the solvent (second row), LNPs (third row), and SNPs (bottom row) for $\phi_{0.10} R_{30}$: (a)--(d), $\phi_{0.10} R_{5}$: (e)--(h), $\phi_{0.10} R_{1}$: (i)--(l), and $\phi_{0.034} R_{5}$: (m)--(p). For each system, each set of curves of the same color corresponds to the snapshot with time indicated in the same color in Fig.~\ref{Figure:snapshots}. The vertical dashed lines indicate the location of the liquid/vapor interface. For clarity, the density profiles for LNPs (SNPs) are shifted upward by $0.1m/\sigma^{3}$ ($0.2m/\sigma^{3}$) successively. The inset in (p) shows a weakly negative gradient of SNP density for $\phi_{0.034}R_{5}$ at time $t = 6 \times 10^5 \tau$ with a dashed horizontal reference line. }
\label{Figure:density}
\end{figure*}

To further quantify how the distributions of the solvent, LNPs, and SNPs evolve during evaporation, we plot their density profiles in Fig.~\ref{Figure:density} at various times corresponding to the snapshots in Fig.~\ref{Figure:snapshots}. The density profiles for the remaining 11 systems can be found in the Supporting Information. The density is defined as $\rho_{i}(z) = n_{i} (z) m_i / (L_x L_y \delta z)$, where $n_{i} (z)$ represents the number of $i$-type particles in the spatial bin $[z - \delta z/2, z + \delta z/2]$ with the bin width $\delta z = 1.0 \sigma$ and $m_i$ is the mass of $i$-type particles. The unit of density is thus $m/\sigma^3$. For a NP occupying several bins, we partition the NP mass to bins based on the partial volume of the NP enclosed by each bin. For computing the solvent density, the solvent particles are treated as point masses and the excluded volume occupied by NPs in each spatial bin is subtracted.\cite{ACSNano2018Note2} To understand the density profiles of the solvent, we also include in Fig.~\ref{Figure:density} (top row) the corresponding local temperature, $T(z)$, which is computed as the mean kinetic energy of solvent beads in the spacial bin $[z-2.5\sigma, z+2.5\sigma]$. The results clearly show evaporative cooling, especially for ultrafast evaporation rates, which leads to negative temperature gradients in the liquid solvent. The calculation of temperature in an evaporating system, which is out of equilibrium, as well as the fact that $T(z)$ has a minimum at the liquid-vapor interface, is discussed in detail in Ref.~\onlinecite{Cheng2011}.

Figure \ref{Figure:density} shows quantitatively the trends that are qualitatively identified from Fig.~\ref{Figure:snapshots}. It is noted that for all the systems at equilibrium there is always a slight density peak near the liquid/vapor interface for the SNPs since they are smaller and their centers can get closer to the interface. When the solvent evaporates very fast, both the LNPs and SNPs are found to accumulate near the interface, as shown in Figs.~\ref{Figure:density}(c) and (d) for $\phi_{0.10} R_{30}$. When the evaporation rate is reduced, the SNPs still accumulate near the interface but the LNPs are almost uniformly distributed in the region below the surface layer where SNPs are concentrated [Figs.~\ref{Figure:density}(g) and (h)]. As a result, the ``small-on-top'' stratification becomes more significant for $\phi_{0.10} R_{5}$. This change is accompanied by a change of the density profile of the solvent as the evaporation rate is reduced. As shown in Fig.~\ref{Figure:density}(b) for $\phi_{0.10} R_{30}$ with a ultrafast evaporation rate, the solvent density increases significantly at the liquid/vapor interface due to strong evaporative cooling [Fig.~\ref{Figure:density}(a)], leading to a large positive gradient of the density profile. The gradients are much smaller for $\phi_{0.10} R_{5}$ with a smaller evaporation rate [Figs.~\ref{Figure:density}(e) and (f)]. In the systems studied here, the NPs are well solvated and form a uniform dispersion in equilibrium prior to solvent evaporation. A density gradient of the solvent that develops during evaporation induces a chemical potential gradient, which drives NPs to regions with a higher solvent density. Our simulations thus indicate a phoretic effect on NP motion when the solvent evaporates ultrafast. Hereafter we use the term ``thermophoresis'' to denote the drift of NPs under a density gradient of the solvent, which is caused by the thermal gradient in the evaporating solvent.\cite{Brenner2011PRE} A similar effect associated with the density gradients of polymers was observed in our previous work where a polymer solution containing NPs underwent drying.\cite{Cheng2016}

As the solvent evaporates, the liquid/vapor interface recedes and the NPs with ${\rm Pe} > 1$ have a tendency to accumulate near the interface. When the SNPs are the major phase and their accumulation at the interface leads to a concentration gradient that is large enough, the LNPs are pushed away by an osmotic pressure induced by the gradient of SNP concentration. This diffusiophoretic mechanism is underlying the current physical models of ``small-on-top'' stratification.\cite{Fortini2016, Zhou2017, Sear2017, Sear2018} However, a positive gradient of solvent density can develop during evaporation because of evaporative cooling of the liquid-vapor interface and its magnitude is large when evaporation is ultrafast. This gradient of solvent density tends to drive all NPs to the interface, but the thermophoretic effect is stronger for LNPs than for SNPs (see the Supporting Information for direct evidence of this behavior). The net effect of the positive gradient of solvent density is thus to push more LNPs toward the interfacial region. The competition between thermophoresis favoring more LNPs near the interface and a fast receding interface, which leads to SNP concentration at the interface and pushes LNPs out of this region via diffusiophoresis, is the key to understand our results. For ultrafast evaporation ($\zeta = 30$), thermophoresis is significant and we observe an accumulation of LNPs just below the skin layer of SNPs, as in the system $\phi_{0.10} R_{30}$ [Figs.~\ref{Figure:density}(c) and (d)]. When the evaporation rate is reduced to $\zeta=5$ as in the system $\phi_{0.10} R_{5}$, the density gradient of the solvent, as well as the temperature gradient, is much smaller [Figs.~\ref{Figure:density}(f) and (e)], which cannot overcome the concentration gradient of SNPs any more in terms of transporting LNPs. Therefore, thermophoresis is strongly suppressed and the LNPs do not accumulate near the interface in $\phi_{0.10} R_{5}$, resulting in stronger ``small-on-top'' stratification [Figs.~\ref{Figure:density}(g) and (h)].

That the solvent density develops a positive gradient in the interfacial region during ultrafast evaporation is due to strong evaporative cooling at the interface,\cite{Cheng2011} as shown in Fig.~\ref{Figure:density} (top row). In this case the diffusion of the solvent toward the interface is driven by a temperature gradient. Evaporative cooling also makes NPs to diffuse more slowly near the receding interface, which increases the P\'{e}clet numbers of those NPs. This effect is stronger for higher evaporation rates which lead to stronger evaporative cooling. The P\'{e}clet numbers used in this paper are defined with the diffusion constants in the solvent at the bulk temperature and are thus the lower bounds of actual values. However, this simplification does not affect the results and conclusions presented in this paper, as discussed in more detail later.

When the evaporation rate is reduced, the degree of interfacial cooling decreases. More examples of the temperature profile at various evaporation rates are included in the Supporting Information. For low evaporation rates, the thermal conduction in the suspension is fast enough to maintain a uniform temperature profile and the density profile of the solvent is almost flat. This situation is realized in the system $\phi_{0.10} R_{1}$ [Figs.~\ref{Figure:density}(i)--(l)], where evaporation is not fast enough to enable SNPs to accumulate at the interface and there is no density gradient of the solvent to drive NPs into the interfacial region either. As a result, the NPs are almost uniformly distributed in the drying film for $\phi_{0.10} R_{1}$.

Comparison of $\phi_{0.10} R_{5}$ with $N_s=19200$ and $\phi_{0.034} R_{5}$ with $N_s=6400$ shows the effect of the initial volume fraction of SNPs, $\phi_s$, on the evaporation-induced stratification. In both cases the evaporation rate is the same ($\zeta = 5$) and the density gradients of the solvent and the temperature gradients are similar [Figs.~\ref{Figure:density}(e), (f), (m), and (n)]. However, the interfacial region in which SNPs are accumulated is wider for $\phi_{0.10} R_{5}$ which has a larger $\phi_s$ [compare Figs.~\ref{Figure:density}(h) and (p)]. In $\phi_{0.10} R_{5}$ the LNPs are almost uniformly distributed in the region below the SNP-rich skin layer [Fig.~\ref{Figure:density}(g)], even though ${\rm Pe}_l \gg 1$. The underlying reason is that the diffusiophoretic force due to the gradient of SNP concentration, which drives the LNPs away from the interface, almost balances the thermophoretic force from the small positive gradient of solvent density, which pushes the LNPs toward the interface. For $\phi_{0.034} R_{5}$ which has a much smaller $\phi_s$, however, there is a strong accumulation of LNPs near the interface [Fig.~\ref{Figure:density}(o)], as $\phi_s$ is too small to yield a noticeable gradient of SNP concentration that is needed to balance the gradient of solvent density. In other words, $\phi_s$ is too small to enable diffusiophoresis to neutralize thermophoresis. In the late stage of drying, the distribution of SNPs in $\phi_{0.034} R_{5}$ even shows a negative gradient and $\rho_s(z)$ decreases slightly toward the interface [Fig.~\ref{Figure:density}(p) and inset], indicating ``large-on-top'' stratification. This trend is qualitatively consistent with the prediction of the existing theories that a transition from ``small-on-top'' to ``large-on-top'' will occur when $\phi_s$ is reduced.\cite{Zhou2017, Sear2017, Sear2018}

\begin{figure}[tb]
\includegraphics[width = 0.4\textwidth]{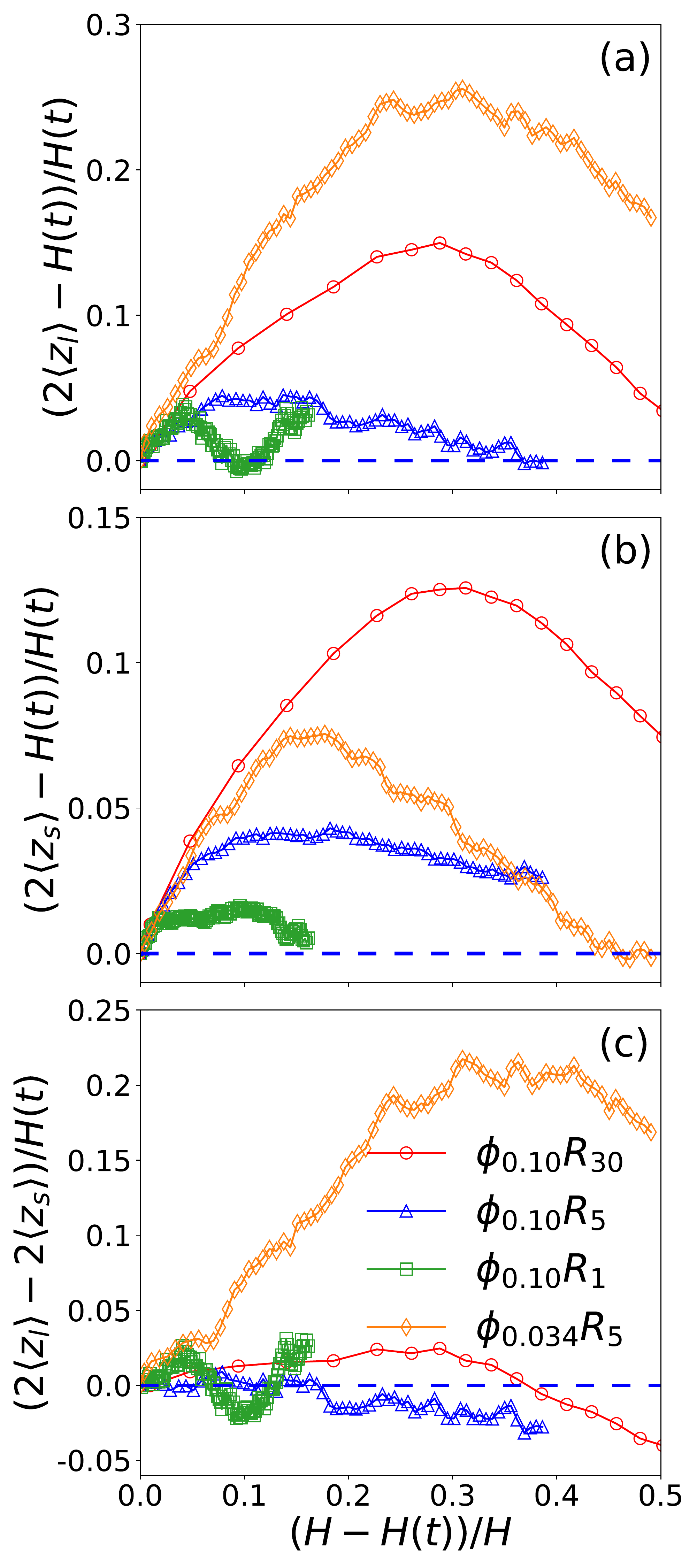}
\caption{Average position in the $z$ direction relative to the center of the film, normalized by $H(t)/2$, is plotted against the extent of drying, quantified as $(H-H(t))/H$, for (a) LNPs and (b) SNPs. Panel (c) shows the average separation between LNPs and SNPs, normalized by $H(t)/2$, as a function of the extent of drying. Data are for $\phi_{0.10} R_{30}$ (red circles), $\phi_{0.10} R_{5}$ (blue triangles), $\phi_{0.10} R_{1}$ (green squares), and $\phi_{0.034} R_{5}$ (orange diamonds).}
\label{Figure:avgz}
\end{figure}

Presently, there is actually no universally adopted criterion on how to identify and quantify stratification. In experiments, especially in those using surface characterization such as AFM measurements, an excess of small particles at the top surface is taken as a signature of ``small-on-top'' stratification as it is difficult to probe depth profiles of particle concentrations.\cite{Makepeace2017, Liu2018} However, this criterion is not suitable for our simulations as an excess of SNPs at the surface of the film even occurs in equilibrium. In theory, a ``small-on-top'' state is usually defined as the one in which large particles have a negative concentration gradient going toward the surface of the film.\cite{Zhou2017, Makepeace2017} Here, to obtain a quantitative measure of the degree of stratification, we use the full concentration profile of NPs and compute the average positions of SNPs and LNPs along the $z$ direction (i.e., normal to the film) as $\langle z_i \rangle= \frac{1}{N_i} \sum\limits_{n=1}^{N_i} z_{in}$ with $i \in \{ l, s\}$, as well as the average separation $ \langle z_l \rangle - \langle z_s\rangle $. The results are shown in Figs.~\ref{Figure:avgz}. At equilibrium, both $ \langle z_l \rangle $ and $ \langle z_s \rangle $ are very close to $H/2$, where $H\equiv H(0)$ is the equilibrium film thickness. If the $i$-type NPs are accumulated (depleted) near the liquid/vapor interface during evaporation, then $\langle z_i \rangle$ becomes larger (smaller) than $H(t)/2$ with $H(t)$ as the film thickness at time $t$. In Figs.~\ref{Figure:avgz}(a) and (b) we plot $ \langle z_l \rangle-H(t)/2 $ and $\langle z_s\rangle -H(t)/2$, all normalized by $H(t)/2$, against $(H-H(t))/H$, which quantifies the extent of drying. Fig.~\ref{Figure:avgz}(a) clearly shows that the LNPs accumulate near the interface for $\phi_{0.10} R_{30}$ (faster evaporation) and $\phi_{0.034} R_{5}$ (smaller $\phi_s$), while they are depleted near the interface in the late stage of drying for $\phi_{0.10} R_{5}$ (reduced evaporation rate, larger $\phi_s$). Fig.~\ref{Figure:avgz}(b) shows in the early stage of drying, the SNPs always accumulate near the receding interface, even for ${\rm Pe}_s \simeq 1$ as in $\phi_{0.10} R_{1}$.

In Fig.~\ref{Figure:avgz}(c) we plot $ \langle z_l \rangle - \langle z_s\rangle $, normalized by $H(t)/2$, as a function of $(H-H(t))/H$. A ``small-on-top'' stratifying state corresponds to $ \langle z_l \rangle - \langle z_s\rangle < 0$ while a ``large-on-top'' case has $ \langle z_l \rangle - \langle z_s\rangle > 0$. A larger negative (positive) value of $ \langle z_l \rangle - \langle z_s\rangle $ indicates stronger small-on-top (large-on-top) stratification. If the distribution of NPs in the film is uniform, then $ \langle z_l \rangle - \langle z_s\rangle \simeq 0$. Our analyses show that the classification scheme adopted here yields results consistent with those based on concentration gradients of particles. To be consistent with the criteria used in the ZJD and SW models,\cite{Zhou2017, Sear2017} we focus on the range of drying up to $H(t)=H/2$ and regard the state at this stage as the stratification outcome. Fig.~\ref{Figure:avgz}(c) shows that ``small-on-top'' only emerges at late times for $\phi_{0.10} R_{30}$ but occurs very quickly for $\phi_{0.10} R_{5}$. It is clear that ``small-on-top'' stratification is enhanced as ${\rm Pe}_s$ is reduced ($\phi_{0.10} R_{30}$$\rightarrow$$\phi_{0.10} R_{5}$). When ${\rm Pe}_s$ is reduced further, a ``small-on-top'' to ``uniform'' transition occurs ($\phi_{0.10} R_{5}$$\rightarrow$$\phi_{0.10} R_{1}$). When $\phi_s$ is reduced at a given ${\rm Pe}_s$, there is a transition from ``small-on-top'' to ``large-on-top''  ($\phi_{0.10} R_{5}$$\rightarrow$$\phi_{0.034} R_{30}$: $\phi_s$ changes from 0.10 to 0.034).

\begin{figure}[bt]
\includegraphics[width = 0.45\textwidth]{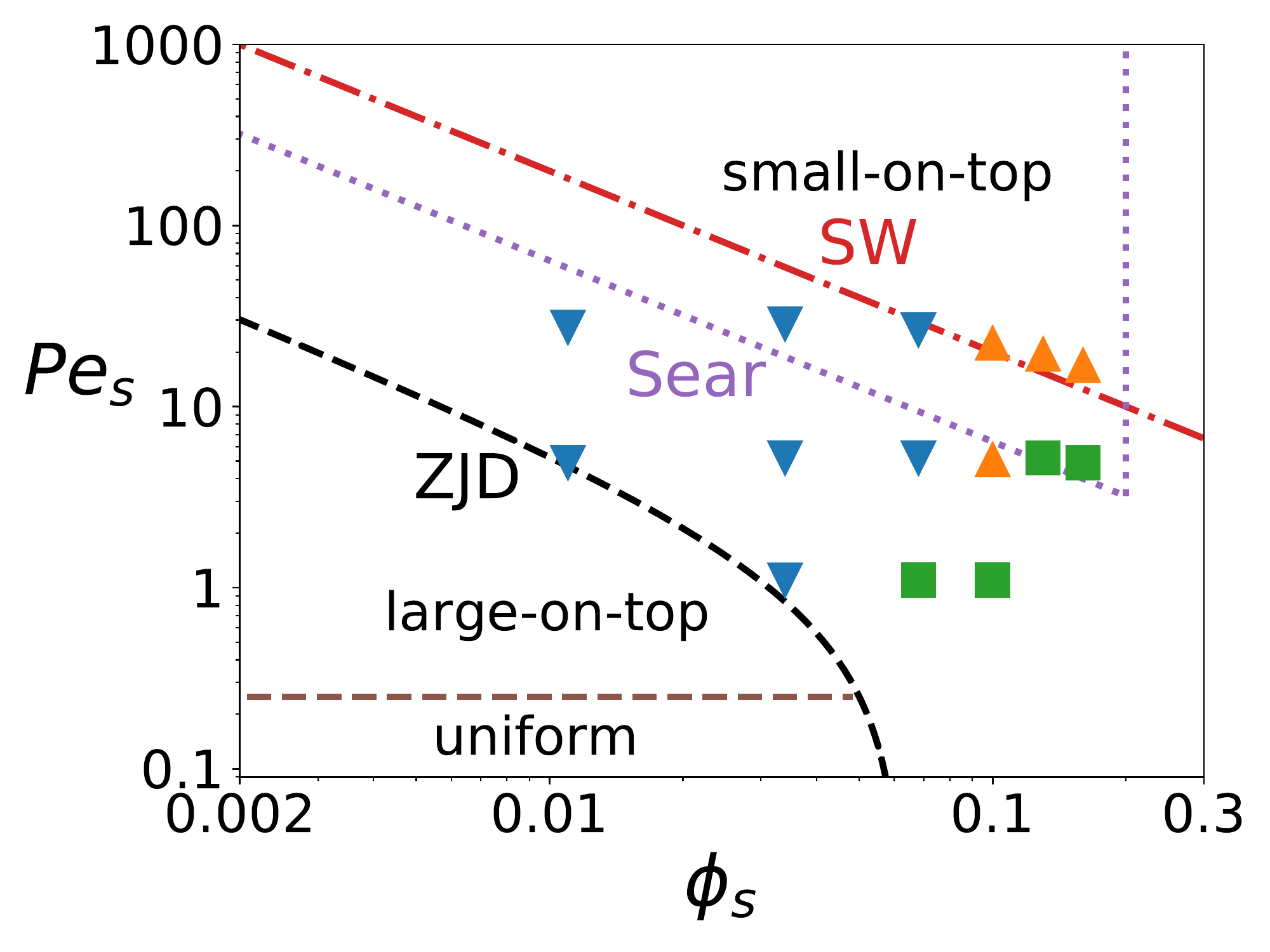}
\caption{The state diagram for the 15 systems studied here with the predictions of the ZJD model,\cite{Zhou2017} the SW model,\cite{Sear2017} and the Sear model.\cite{Sear2018} The systems showing ``small-on-top'' or ``large-on-top'' stratification are indicated by upward or downward triangles, respectively. The systems that do not show stratified distributions of NPs are designated as ``uniform'' and indicated with squares. The ZJD model predicts that ``small-on-top'' occurs when ${\rm Pe}_s \gtrsim 1/(\alpha^2\phi_s)-1$ (black dashed line) and ``uniform'' occurs when ${\rm Pe}_s < 1/\alpha$ (brown dashed line). The SW model predicts that ``small-on-top'' occurs when ${\rm Pe}_s \gtrsim 2/\phi_s$ (red dash-dotted line). The Sear model predicts that ``small-on-top'' occurs only for a finite range of $\phi_s$, corresponding to $0.64/{\rm Pe}_s < \phi_s < 0.20$ (purple dotted line).}
\label{Figure:state_diagram}
\end{figure}

In Fig.~\ref{Figure:state_diagram} all 15 systems studied here are included in the state diagram in the ${\rm Pe}_s$--$\phi_s$ plane, and compared to the predictions of the ZJD model,\cite{Zhou2017} the SW model,\cite{Sear2017} and the Sear model.\cite{Sear2018} The 4 systems shown in Figs.\ref{Figure:snapshots}--\ref{Figure:avgz} are already classified. The identification of the stratifying outcome for the remaining 11 systems is included in the Supporting Information. In our simulations, only 4 systems show ``small-on-top'' stratification, including $\phi_{0.10} R_{30}$, $\phi_{0.10} R_{5}$, $\phi_{0.13} R_{30}$, and $\phi_{0.16} R_{20}$. Other 4 systems have a uniform distribution of NPs after drying, including $\phi_{0.068} R_{1}$, $\phi_{0.10} R_{1}$, $\phi_{0.13} R_{5}$, and $\phi_{0.16} R_{5}$. The remaining 7 exhibit ``large-on-top'' stratification. Note that all systems simulated here are in the ``small-on-top'' regime predicted by the ZJD model. However, the simulation data show that ``small-on-top'' stratification only occurs at ${\rm Pe}_s$ and $\phi_s$ much higher than the threshold value from the ZJD model that predicts $\alpha^2({\rm Pe}_s +1)\phi_s>1$. The comparison thus indicates that the ZJD model significantly overestimates ``small-on-top'' stratification, in agreement with Makepeace \textit{et al.} \cite{Makepeace2017} and Sear and Warren \cite{Sear2017}. 

As discussed before, the P\'{e}clet numbers used to construct Fig.~\ref{Figure:state_diagram} are the lower bounds. If the temperature dependence of the diffusion constants were accounted for, then the actual P\'{e}clet numbers might even be higher especially for high evaporation rates (i.e., for $\zeta = 30$ or 5), shifting the corresponding data points in Fig.~\ref{Figure:state_diagram} upward. Furthermore, the amount of shift is very small as the P\'{e}clet numbers enter Fig.~\ref{Figure:state_diagram} on a logarithmic scale. As a result, Fig.~\ref{Figure:state_diagram} and the discussion below are not affected by the simplification adopted here that the P\'{e}clet numbers are defined using the diffusion constants in the solvent thermalized at $T = 1.0\epsilon/k_{B}$ (see Methods section for more details) and ignoring their potential variation during evaporation because of evaporative cooling.

It is challenging to distinguish the SW model and the Sear model using our simulation results. The SW model predicts that the ``small-on-top'' regime roughly corresponds to ${\rm Pe}_s \gtrsim 2/\phi_s$.\cite{Sear2017} Our data fit to this prediction reasonably well with 3 systems exhibiting ``small-on-top'' as expected by the SW model. This agreement indicates that the explicit solvent model used here has successfully captured the back-flow of the solvent when NPs drift, as emphasized in the SW model. This back-flow largely cancels out the osmotic pressure on LNPs from the concentration gradient of SNPs. As a result, much larger ${\rm Pe}_s$ and $\phi_s$ are needed to drive a system into the ``small-on-top'' regime. However, our simulations also indicate that the solvent develops negative temperature and positive density gradients for ultrafast evaporation because of evaporative cooling at the interface. The presence of thermophoretic effects associated with these density gradients can explain the deviation of the simulation results from the prediction of the SW model. For example, the system $\phi_{0.10}R_5$ is predicted to be in the ``large-on-top'' regime but actually shows ``small-on-top'' stratification, which as discussed earlier is due to thermophoresis of NPs from the density gradient of the solvent. As another example, the systems $\phi_{0.068} R_{30}$ is at the boundary of the ``small-on-top'' regime according to the SW model. However, for this ultrafast evaporating system the large positive gradient of solvent density pushes LNPs toward the interface much more strongly than SNPs, i.e., thermophoresis is significant. Consequently, $\phi_{0.068} R_{30}$ shows clear ``large-on-top'' stratification.

Our data also seem to be roughly consistent with the Sear model,\cite{Sear2018} which predicts that ``small-on-top'' stratification only occurs when $0.64/{\rm Pe}_s < \phi_s < 0.2$. Fig.~\ref{Figure:state_diagram} shows that the 4 ``small-on-top'' systems identified in our simulations are roughly consistent with this prediction. However, $\phi_{0.068} R_{30}$ and $\phi_{0.034} R_{30}$ are in the ``small-on-top'' regime predicted by the Sear model, but actually are identified as ``large-on-top'' in our simulations. In these systems where $\phi_s$ is small, the thermophoretic effects from the density gradients of the solvent dominate, which push LNPs toward the interface strongly and drive the systems into the ``large-on-top'' regime. 

Both the SW and Sear models predict that the boundary of the ``small-on-top'' regime is roughly at ${\rm Pe}_s \phi_s \gtrsim c$ with $c$ at the order of 1.\cite{Sear2017,Sear2018} This prediction is supported by our simulation results. To test the Sear model further, one would need data for $\phi_s > 0.2$. However, with the present model, if $\phi_s$ is too large, some SNPs move into the vapor during evaporation.\cite{ACSNano2018Note3} At present, the regime of very high volume fractions of SNPs remains an open problem.

Limitations of the computational model used here should be noted. The SNPs in this study have a diameter about 5 times of the size of the solvent particle. If we denote the P\'{e}clet number of the solvent as ${\rm Pe}_0$, then ${\rm Pe}_0 \simeq {\rm Pe}_s/5$. In our simulations, ${\rm Pe}_s$ varies from about 1 to 30. As a result, ${\rm Pe}_0$ is about 0.2 to 6. In most experiments, ${\rm Pe}_0$ is much less than 0.1. This comparison indicates our simulations are all in the ultrafast evaporation regime from an experimental perspective.

Similar conclusions can be drawn if we examine the receding speed of the liquid/vapor interface, $v$. The lowest value of $v$ in our simulations is $\sim 4\times 10^{-5}\sigma/\tau$. With a typical value of $\sigma/\tau$ at 100 m$/$s, this speed is about 4 mm$/$s in simulations. For water evaporating under ambient conditions, $v$ is typically about 0.1 $\mu$m$/$s. Recently, Utgenannt \textit{et al.} used infrared radiation to speed up the evaporation of water and increased $v$ to about 2 $\mu$m$/$s.\cite{Utgenannt2016}. In the experiment of Luo \textit{et al.}, $v$ is about 2.5 $\mu$m$/$s.\cite{Luo2008} Even so, the value of $v$ in our simulations is still about $2\times 10^3$ times larger than that in the experiments. This large factor can be understood as follows. In a typical experiment on the drying of particle suspensions, the thickness of films is usually around 0.1 to 1 mm. However, in our simulations, the film thickness is about $300\sigma$, or 150 nm if we set $\sigma = 0.5$ nm. To achieve the same P\'{e}clet number, ${\rm Pe} \equiv vH/D$, the value of $v$ in our simulations has to be larger than that in a typical experiment by a factor of about $10^3$ to $10^4$. However, if a drying experiment was performed on a liquid film with submicron thickness containing NPs, the evaporation rates (i.e., the values of $v$) would have to be similar to those studied here in order to drive the system into the regime where ``small-on-top'' stratification might occur. Density/temperature gradients are expected to develop in such liquid films that undergo ultrafast drying and the thermophoretic effects found in our simulations may become experimentally relevant. 

The solvent in this study is modeled as a LJ liquid at temperature $T=1.0\epsilon/k_{\rm B}$, where $\epsilon$ is the unit of energy (see the Methods section). This temperature is about $0.9T_c$, where $T_c$ is the critical temperature of a LJ liquid. At this temperature the solvent can evaporate extremely fast, which leads to strong evaporative cooling and large density gradients near the interface. For water with $T_c = 647$ K, this condition would correspond to a temperature around 600 K and a pressure around 120 atm to maintain a liquid-vapor coexistence at this temperature. If water evaporates at 600 K into a vacuum, then the evaporation rate and the corresponding receding speed of the interface will be comparable to those in our simulations. The density gradients of the liquid are also expected to emerge in such systems.

In order to design MD simulations of the drying of NP suspensions that are more comparable to typical experiments, one would need to decrease $v$, but keep ${\rm Pe}_s\sim ~ 1$. One viable possibility is to decrease the diffusion constant of NPs. For example, larger particles can be used but they would require more solvent particles to form suspensions, rendering very big systems that may be inaccessible with current computational resources. Another way is to make the solvent more viscous with regard to the diffusion of NPs but to maintain the liquid-vapor coexistence, which is needed for the evaporation process to be fast enough to be modeled via MD. Several options include tuning the NP-solvent interactions to slow down the diffusion of NPs or adding other solutes such as polymer chains into the suspension to increase its viscosity. The NP-NP interactions are another factor that may be explored. In this paper to be consistent with most theoretical models based on hard spheres, we set the direct NP-NP interactions to be purely repulsive, though there exist weak solvent-mediated attractions between NPs. It is interesting to see how the outcome of drying changes when NPs strongly attract each other. All these remain potential directions for future studies.

\bigskip
\noindent{\bf CONCLUSIONS}

\noindent MD simulations with an explicit solvent are reported on the drying of bidisperse NP suspensions and indicate that ``small-on-top'' stratification occurs when the evaporation rate (quantified by the P\'{e}clet numbers of the NPs: ${\rm Pe}_l$ and ${\rm Pe}_s$) and the volume fraction of the smaller particles ($\phi_s$) are large enough. Boundary of the ``small-on-top'' regime is found to be roughly ${\rm Pe}_s \phi_s \gtrsim c$ with $c \sim 1$, consistent with the SW model ($c=2$) \cite{Sear2017} and the Sear model ($c=0.64$) \cite{Sear2018}. In the ${\rm Pe}_s$--$\phi_s$ plane, this boundary is to the right of and above the boundary, roughly ${\rm Pe}_s \phi_s \gtrsim \alpha^{-2}$ for small $\phi_s$, predicted by the ZJD model that treats the solvent as an implicit viscous background.\cite{Zhou2017} The two predictions differ roughly by a factor of $\alpha^2$, which can be quite large if the particle size ratio $\alpha \gg 1$. As pointed out by Sear and Warren,\cite{Sear2017} this is due to the fact that the implicit solvent model neglects the back-flow of the solvent when particles drift, which largely cancels out the diffusiophoretic drift of LNPs induced by the concentration gradient of SNPs. As a result, the drift velocity of LNPs in a drying film is overestimated by a factor of $\alpha^2$ by the implicit solvent model.\cite{Fortini2016,Zhou2017} Our results are consist with the SW and Sear models, confirming that it is important to include the solvent explicitly in a physical model of stratification, or generally the drift of particles, in a suspension.

Our simulations further reveal that the solvent can develop positive density gradients in ultrafast evaporating suspensions because of evaporative cooling of the interface, which leads to thermophoretic effects on particle motion. For a bidisperse NP suspension undergoing quick drying (${\rm Pe}_l > {\rm Pe}_s\gg 1$), the net thermophoretic effect is to push more LNPs toward the interfacial region. This effect can lead to ``large-on-top'' stratification at high ${\rm Pe}_s$ even when ``small-on-top'' stratification is expected. This deviation is due to thermophoresis which favors ``large-on-top'' and competes with a fast receding interface that drives ``small-on-top'' as emphasized in the diffusiophoretic models.\cite{Fortini2016,Zhou2017,Sear2017,Sear2018} Similar effect can also make ``small-on-top'' stratification stronger as the evaporation rate is reduced, since the thermophoretic driving which favors LNPs on top is mitigated. Our results confirm the necessity of considering solvent explicitly in theory and modeling. In the presence of gravity, a convective flow can form to balance the solvent gradient from ultrafast evaporation. This points to the potential need of considering convective flow in next-generation physical models of stratifying phenomena.

Because of thermophoresis that drives more LNPs toward the receding interface, the simulations reported here only show weak ``small-on-top'' stratification when it actually occurs. In other cases thermophoresis is strong enough that the stratification of large and small particles is reversed to ``large-on-top'' even when ``small-on-top'' is predicted by the diffusiophoretic models.\cite{Sear2017,Sear2018} In order to promote ``small-on-top'' stratification, thermophoresis needs to be suppressed. Indeed, we have observed stronger ``small-on-top'' stratification if all the liquid and vapor are thermalized at a constant temperature during evaporation, where thermal and density gradients and associated thermophoretic transport are removed. However, thermophoresis can also be exploited to produce ``large-on-top'' stratification under circumstances where it is not expected. Our results thus indicate that phoretic effects can be used as a knob to control the outcome of stratification. Work along this line is in progress and will be reported in the future.

\bigskip
\noindent{\bf METHODS}

\noindent The solvent is modeled by beads with a mass $m$ interacting through a standard Lennard-Jones (LJ) potential, $U_{\text{LJ}}(r) = 4\epsilon \left[ (\sigma/r)^{12} - (\sigma/r)^6 - (\sigma/r_c)^{12} + (\sigma/r_c)^6 \right]$, where $r$ is the distance between the centers of two beads, $\epsilon$ the unit of energy, and $\sigma$ the diameter of beads. The interaction is truncated at $r_c = 3.0 \sigma$. The nanoparticles (NPs) are modeled as a uniform distribution of LJ particles of a mass density $1.0m/\sigma^3$. The diameter of a large nanoparticle (LNP) is $d_l = 20\sigma$ and of a small nanoparticle (SNP) is $d_s = 5 \sigma$. The mass is $m_l = 4188.8m$ and $m_s = 65.4m$, respectively, for LNPs and SNPs. The NP-NP interaction potential can be determined analytically by integrating over all the interacting LJ particles within the two NPs.\cite{Everaers2003, IntVeld2008} The same Hamaker constant $A_{ns} = 39.48\epsilon$ sets the strength of interaction between all the NPs. To avoid flocculation,\cite{IntVeld2009, Grest2011} we set the NP-NP interactions to be purely repulsive by truncating the potential at $20.574\sigma$, $13.086\sigma$, and $5.595\sigma$, respectively, for the LNP/LNP, LNP/SNP, and SNP/SNP pairs. The interaction between a solvent bead and a NP can be described similarly with an integrated LJ potential. We set the interaction strength between the solvent and the NPs as $A_{ns} = 100\epsilon$ and truncate the potential at $d/2+4\sigma$ where $d$ is the NP diameter. As a result, both the LNPs and SNPs are fully solvated by the solvent.\cite{Cheng2012}

All the solvent beads are placed in a rectangular simulation box with dimensions $L_x \times L_y \times L_z$, where $L_x = 201\sigma$, $L_y = 201\sigma$, and $L_z = 477\sigma$. These beads form a liquid film with a thickness $H\sim 300\sigma$ (see Table \ref{Table:system} of the main text for the value of $H$ in each system), which serves as the solvent, and a vapor phase above it. Periodic boundary conditions are imposed in the $x$-$y$ plane, in which the liquid/vapor interface is located. The NPs are randomly dispersed in the liquid solvent and the system is equilibrated before evaporation is turned on. All the particles are confined between two flat walls at $z=0$ and $z=L_z$ via a LJ 9-3 potential, $U_W (h) = \epsilon_W \left[ (2/15)(D/h)^9 - (D/h)^3 - (2/15)(D/h_c)^9 + (D/h_c)^3 \right]$, where $\epsilon_W = 2.0\epsilon$ is the interaction strength, $D$ the characteristic length, $h$ the distance between the center of the particle and the wall, and $h_c$ the cutoff. We set $D=\sigma$ and $h_c = 3\sigma$ ($0.8583\sigma$) at the lower (upper) wall for the solvent/wall interactions; the upper wall is thus repulsive for the solvent. For the NP/wall interactions we set $D = d/2$ and $h_c=0.8583D$ at both walls to make them purely repulsive for all the NPs.

All simulations were performed with the Large-scale Atomic/Molecular Massively Parallel Simulator (LAMMPS).\cite{Plimpton1995} The equations of motion were integrated using a velocity-Verlet algorithm with a time step $\delta t = 0.01 \tau$, where $\tau = \sigma (m/\epsilon)^{1/2}$ is the time unit. During the equilibration, Langevin dynamics were applied to all the particles with a damping constant $\Gamma = 0.1\tau^{-1}$ at a reduced temperature $T = 1.0\epsilon/k_{B}$. We equilibrated the system for at least $4 \times 10^5 \tau$ so that all the NPs were well dispersed in the solvent. In the evaporation runs, the Langevin thermostat was applied only for the solvent and the NPs within $10\sigma$ of the lower wall.\cite{Cheng2011} The evaporation was implemented by removing the vapor beads in the deletion zone $[L_z -100\sigma, L_z]$, which was about $70\sigma$ away from the equilibrium liquid/vapor interface. The evaporation rate was controlled by varying the rate at which the vapor beads in the deletion zone were removed from the system. As the initial thickness of the liquid film is about $300\sigma$ and we only focus on the range of drying where the film is still more than half of its initial thickness, the distance from the thermalized layer to the liquid-vapor interface is thus at least $140\sigma$. This separation is more than sufficient to ensure that the evaporating behavior at the interface is not affected by the thermostat employed in our simulations.\cite{Cheng2011,Heinen2016}

\section*{Acknowledgment}

Acknowledgment is made to the Donors of the American Chemical Society Petroleum Research Fund (PRF \#56103-DNI6), for support of this research. This work was performed, in part, at the Center for Integrated Nanotechnologies, an Office of Science User Facility operated for the U.S. Department of Energy (DOE) Office of Science. Sandia National Laboratories is a multimission laboratory managed and operated by National Technology and Engineering Solutions of Sandia, LLC., a wholly owned subsidiary of Honeywell International, Inc., for the U.S. Department of Energy's National Nuclear Security Administration under contract DE-NA-0003525.

%\bibliography{evap}

\clearpage
\onecolumngrid
\renewcommand{\thefigure}{E\arabic{figure}}
\setcounter{figure}{0}    
\renewcommand{\thepage}{SI-\arabic{page}}
\setcounter{page}{1}    
\begin{center}
{\bf SUPPORTING INFORMATION}
\end{center}

%\foreach \x in {1,...,22}
%{%
%\clearpage
%\includepdf[pages={\x}]{stratification_SI_04-01-2018.pdf}
%}

%\bigskip
\noindent {\bf ADDITIONAL RESULTS AND DISCUSSION}
%\bigskip

\begin{figure}[tp]
\includegraphics[width = 0.75\textwidth]{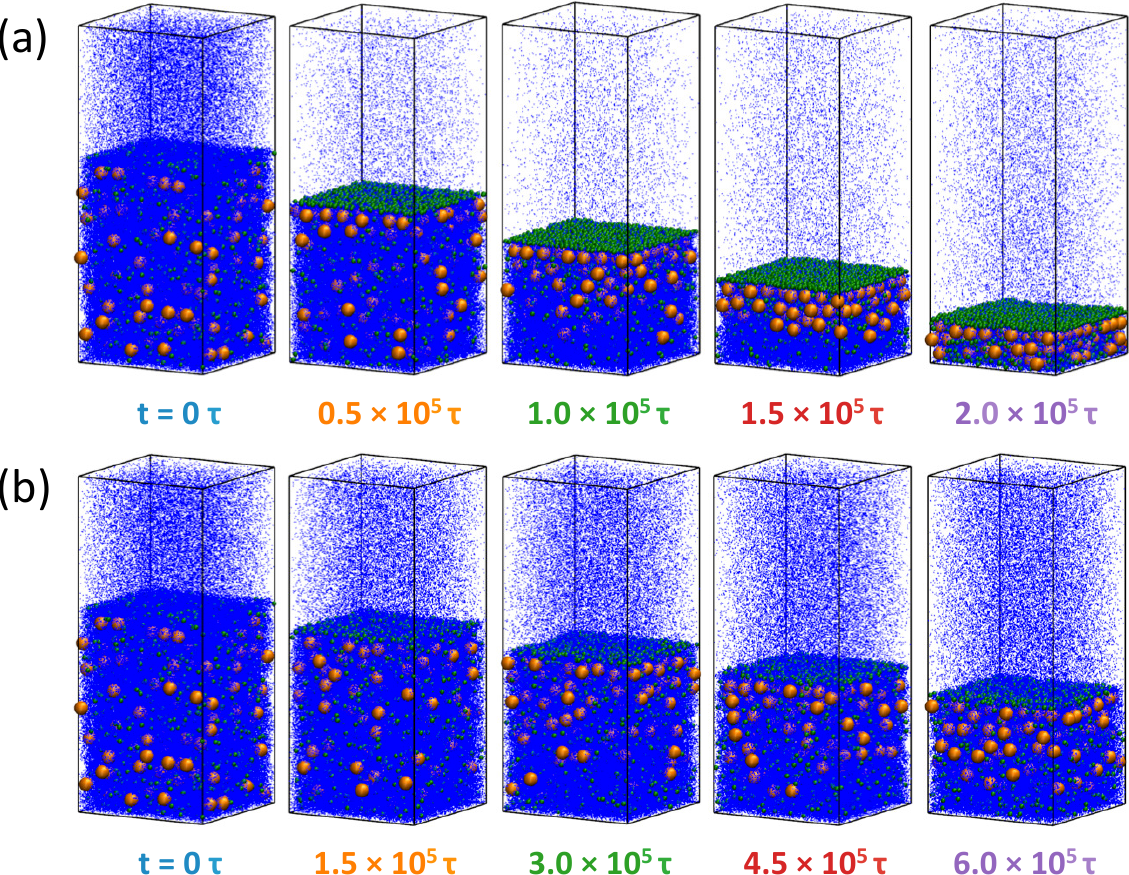}
\caption{Snapshots of (a) $\phi_{0.011}R_{30}$ and (b) $\phi_{0.011}R_{5}$ with $N_s = 1920$. Color code: LNPs (orange), SNPs (green), and solvent (blue). For clarity, only 5\% of the solvent beads are visualized. In the last frame the volume fractions of nanoparticles are: (a) $\phi_l$ = 0.34, $\phi_s$ = 0.051;(b) $\phi_l$ = 0.12, $\phi_s$ = 0.018.}
\label{Figure:1920SNP_snapshots}
\end{figure}

\begin{figure}[tp]
\includegraphics[width = 0.75\textwidth]{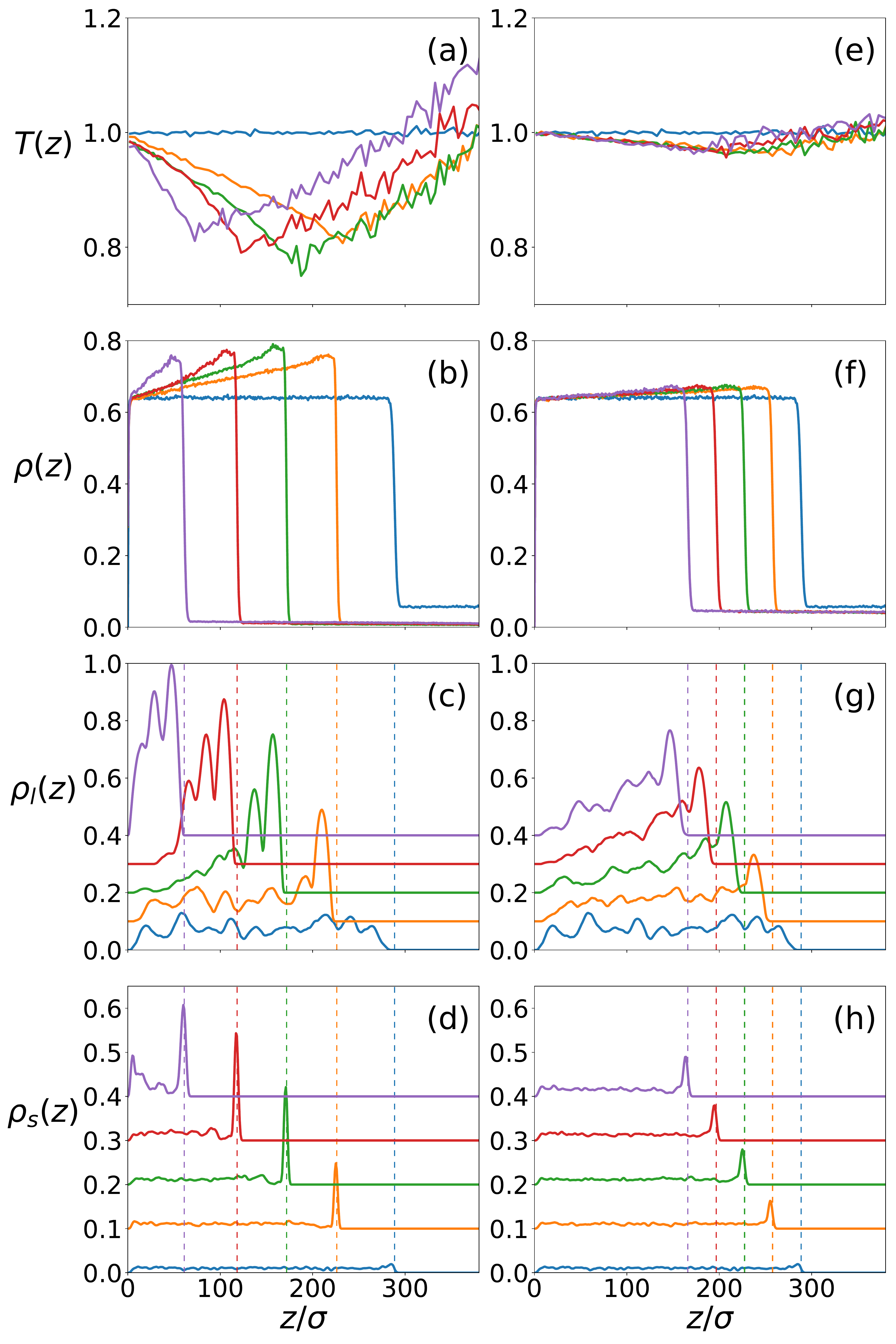}
\caption{Temperature and density profiles for $\phi_{0.011}R_{30}$ (left column) and $\phi_{0.011}R_{5}$ (right column): (a) and (e) temperature; (b) and (f) solvent; (c) and (g) LNPs; (d) and (h) SNPs. The vertical dashed lines indicate the location of the liquid/vapor interface. For clarity, the density profiles for NPs are shifted upward successively by $0.1m/\sigma^{3}$.}
\label{Figure:1920SNP_density}
\end{figure}

\begin{figure}[tp]
\includegraphics[width = 0.5\textwidth]{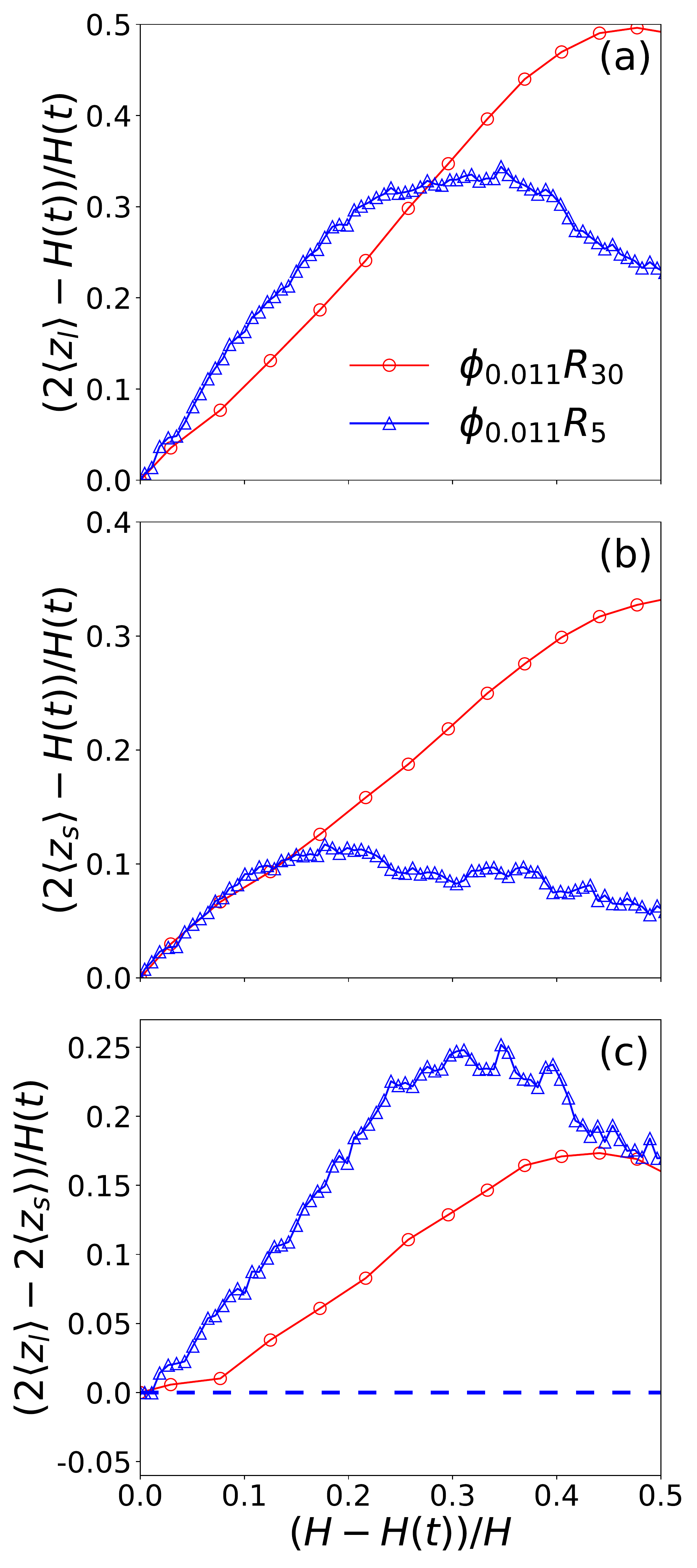}
\caption{Average position in the $z$ direction relative to the center of the film, normalized by $H(t)/2$, is plotted against time $(H-H(t))/H$ for (a) LNPs and (b) SNPs. Panel (c) shows the average separation between LNPs and SNPs, normalized by $H(t)/2$, as a function of $(H-H(t))/H$. Data are for system $\phi_{0.011}R_{30}$ (red circles) and $\phi_{0.011}R_{5}$ (blue triangles).}
\label{Figure:1920SNP_order}
\end{figure}

\noindent {\bf 1920-SNP Systems:} Figure~\ref{Figure:1920SNP_snapshots} shows snapshots of $\phi_{0.011}R_{30}$ and $\phi_{0.011}R_{5}$ with $N_s=1920$ at various times during evaporation. The corresponding temperature and density profiles are shown in Fig.~\ref{Figure:1920SNP_density}. For these two systems, the initial volume fraction of the smaller nanoparticles (SNPs) is $\phi_s = 0.011\simeq \phi_l/6$. During solvent evaporation, a skin layer of SNPs forms in both systems but is denser in $\phi_{0.011}R_{30}$, which has a higher evaporation rate. The accumulation of larger nanoparticles (LNPs) below the skin layer of SNPs is obvious for both systems and more significant for $\phi_{0.011}R_{30}$. Fig.~\ref{Figure:1920SNP_order} shows the average positions of LNPs and SNPs ($ \langle z_l \rangle $ and $ \langle z_s\rangle $) and their mean separation ($ \langle z_l \rangle - \langle z_s\rangle $) against $(H-H(t))/H$, where $H(t)$ is the film thickness at time $t$ and $H\equiv H(0)$ is the equilibrium film thickness prior to evaporation. The quantity $(H-H(t))/H$ defines the extent of drying. The mean separation, $ \langle z_l \rangle - \langle z_s\rangle $, characterizes the state of stratification with $ \langle z_l \rangle - \langle z_s\rangle > 0$ signals ``large-on-top'' while $ \langle z_l \rangle - \langle z_s\rangle < 0$ indicates ``small-on-top''. From Fig.~\ref{Figure:1920SNP_order}, it is clear that both $\phi_{0.011}R_{30}$ and $\phi_{0.011}R_{5}$ show the ``large-on-top'' stratification. This classification is consistent with the density profiles of NPs shown in Fig.~\ref{Figure:1920SNP_density}, where the density profile of LNPs exhibits a positive gradient approaching the interfacial region. Though there is a surface accumulation of SNPs, the density profile of SNPs shows a negative gradient in the central region of the film.

\begin{figure}[tp]
\includegraphics[width = 0.75\textwidth]{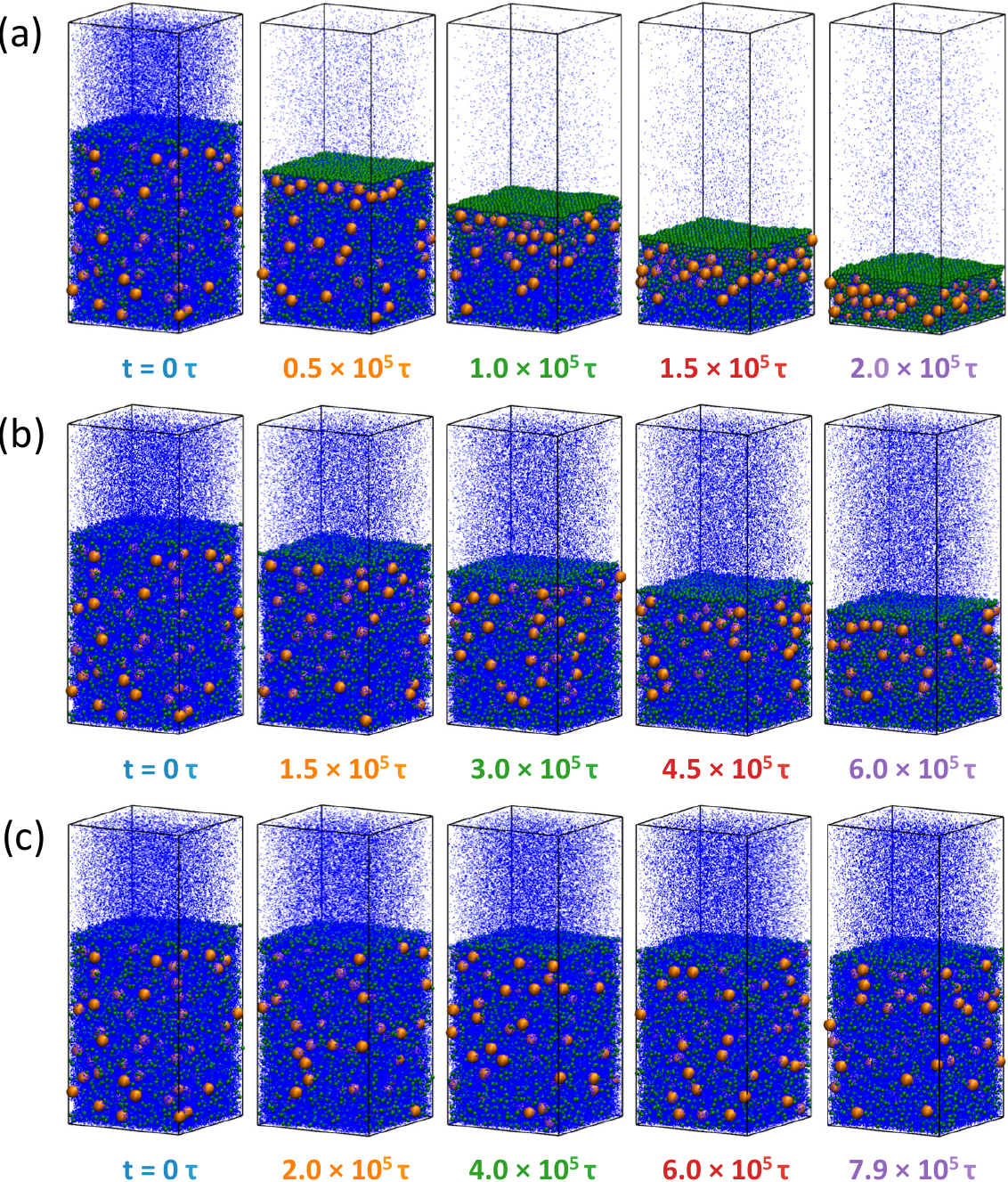}
\caption{Snapshots of (a) $\phi_{0.034}R_{30}$, (b) $\phi_{0.034}R_{5}$, and (c) $\phi_{0.034}R_{1}$ with $N_s = 6400$. Color code: LNPs (orange), SNPs (green), and solvent (blue). For clarity, only 5\% of the solvent beads are visualized. In the last frame the volume fractions of nanoparticles are: (a) $\phi_l$ = 0.26, $\phi_s$ = 0.13;(b) $\phi_l$ = 0.11, $\phi_s$ = 0.057; (c) $\phi_l$ = 0.077, $\phi_s$ = 0.038.}
\label{Figure:6400SNP_snapshots}
\end{figure}

\begin{figure}[tp]
\includegraphics[width = \textwidth]{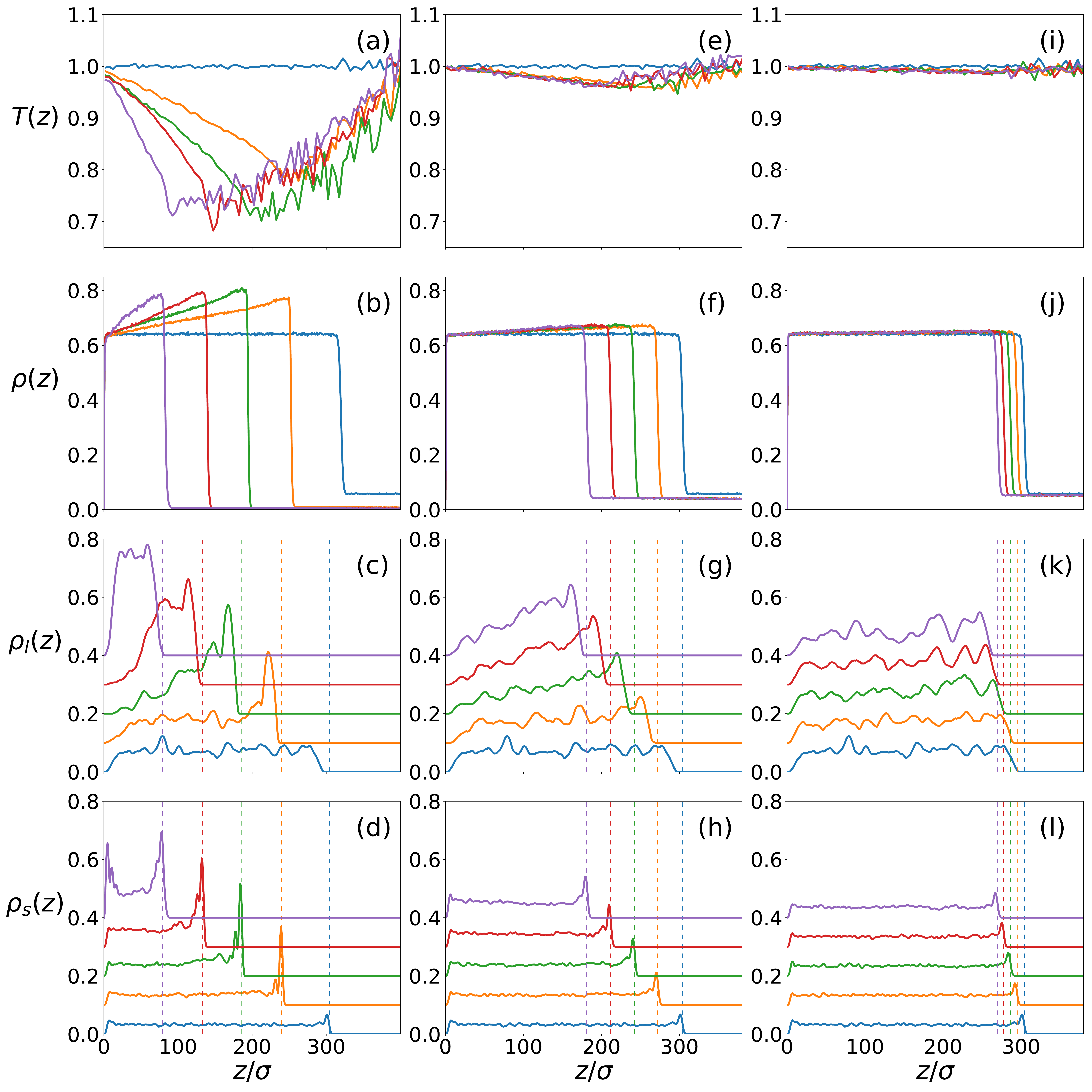}
\caption{Temperature and density profiles for $\phi_{0.034}R_{30}$ (left column), $\phi_{0.034}R_{5}$ (middle column), and $\phi_{0.034}R_{1}$ (right column): temperature (top row); solvent (second row); LNPs (third row); SNPs (bottom row). The vertical dashed lines indicate the location of the liquid/vapor interface. For clarity, the density profiles for NPs are shifted upward successively by $0.1m/\sigma^{3}$.}
\label{Figure:6400SNP_density}
\end{figure}

\begin{figure}[tp]
\includegraphics[width = 0.5\textwidth]{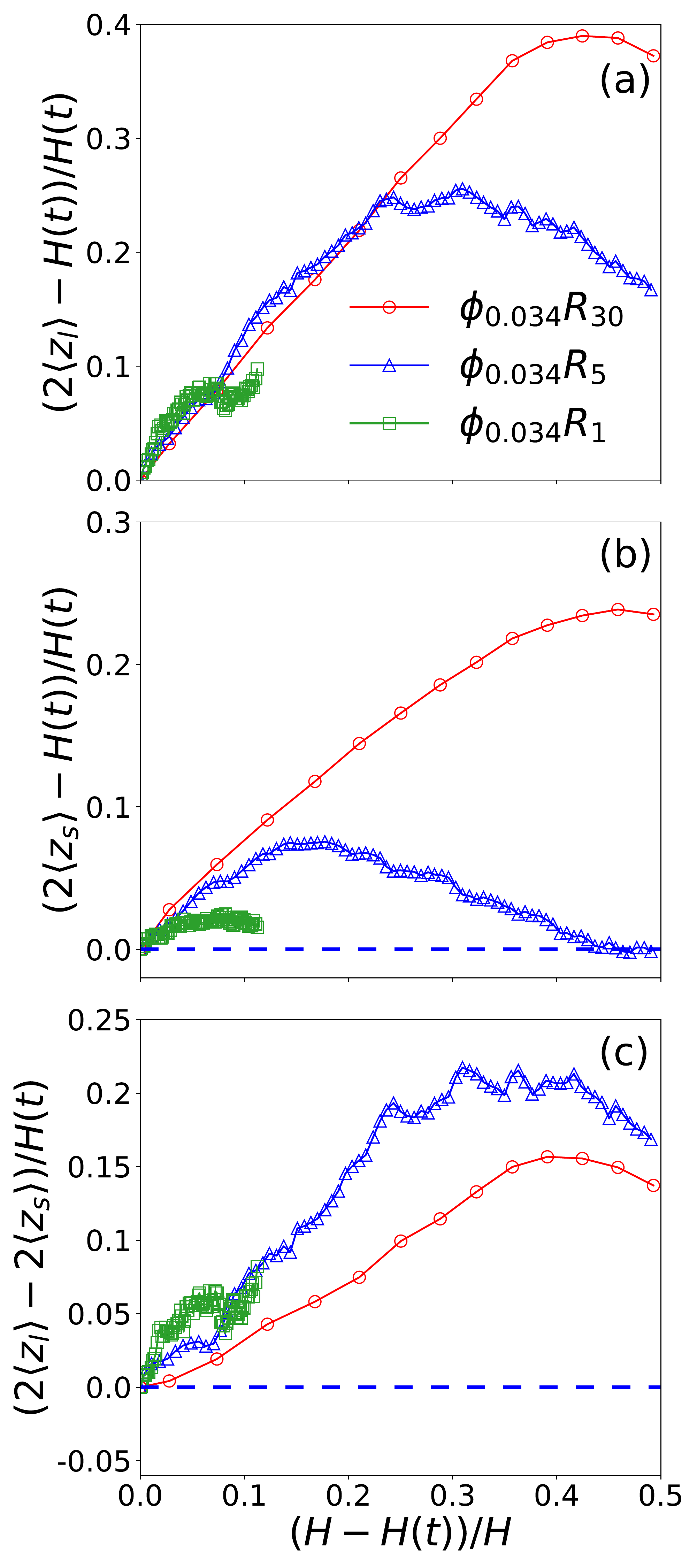}
\caption{Average position in the $z$ direction relative to the center of the film, normalized by $H(t)/2$, is plotted against time $(H-H(t))/H$ for (a) LNPs and (b) SNPs. Panel (c) shows the average separation between LNPs and SNPs, normalized by $H(t)/2$, as a function of $(H-H(t))/H$. Data are for system $\phi_{0.034}R_{30}$ (red circles), $\phi_{0.034}R_{5}$ (blue triangles) and $\phi_{0.034}R_{1}$ (green squares).}
\label{Figure:6400SNP_order}
\end{figure}

\noindent {\bf 6400-SNP Systems:} Figure~\ref{Figure:6400SNP_snapshots} shows snapshots of $\phi_{0.034}R_{30}$, $\phi_{0.034}R_{5}$, and $\phi_{0.034}R_{1}$ with $N_s=6400$ at various times during evaporation. The corresponding temperature and density profiles are shown in Fig.~\ref{Figure:6400SNP_density} and the average positions of NPs and their mean separation are shown in Fig.~\ref{Figure:6400SNP_order}. All these 3 systems behave qualitatively similar as the 2 systems with $N_s=1920$. This is expected as for the 3 systems, $\phi_s=0.034$, which is still only about $1/2$ of $\phi_l$. Based on the density profiles and the order parameters, we classify these 3 systems as ``large-on-top'' as well.

\begin{figure}[tp]
\includegraphics[width = 0.75\textwidth]{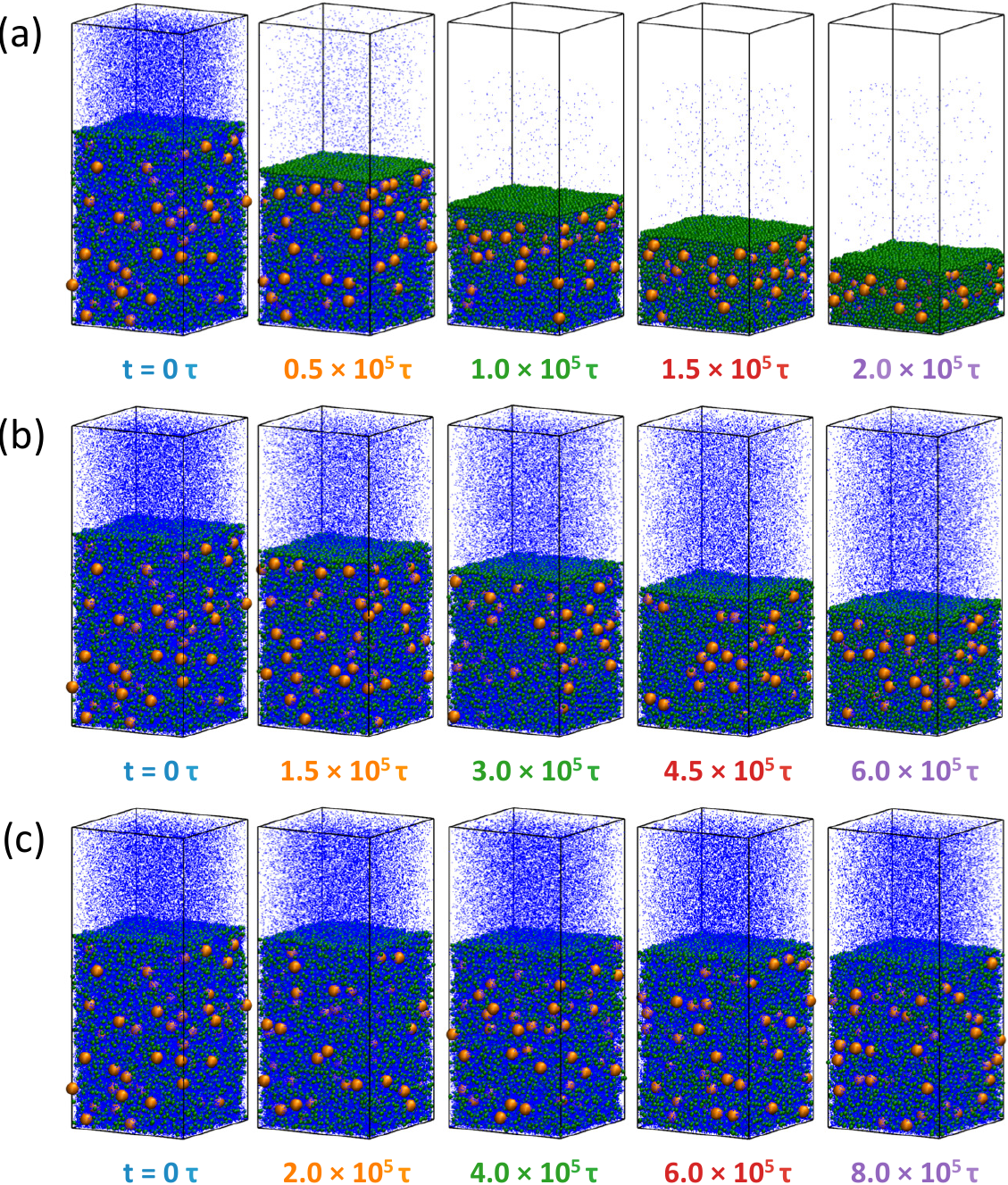}
\caption{Snapshots of (a) $\phi_{0.068}R_{30}$, (b) $\phi_{0.068}R_{5}$, and (c) $\phi_{0.068}R_{1}$ with $N_s = 12800$. Color code: LNPs (orange), SNPs (green), and solvent (blue). For clarity, only 5\% of the solvent beads are visualized. In the last frame the volume fractions of nanoparticles are: (a) $\phi_l$ = 0.21, $\phi_s$ = 0.21;(b) $\phi_l$ = 0.11, $\phi_s$ = 0.11; (c) $\phi_l$ = 0.076, $\phi_s$ = 0.076.}
\label{Figure:12800SNP_snapshots}
\end{figure}

\begin{figure}[tp]
\includegraphics[width = \textwidth]{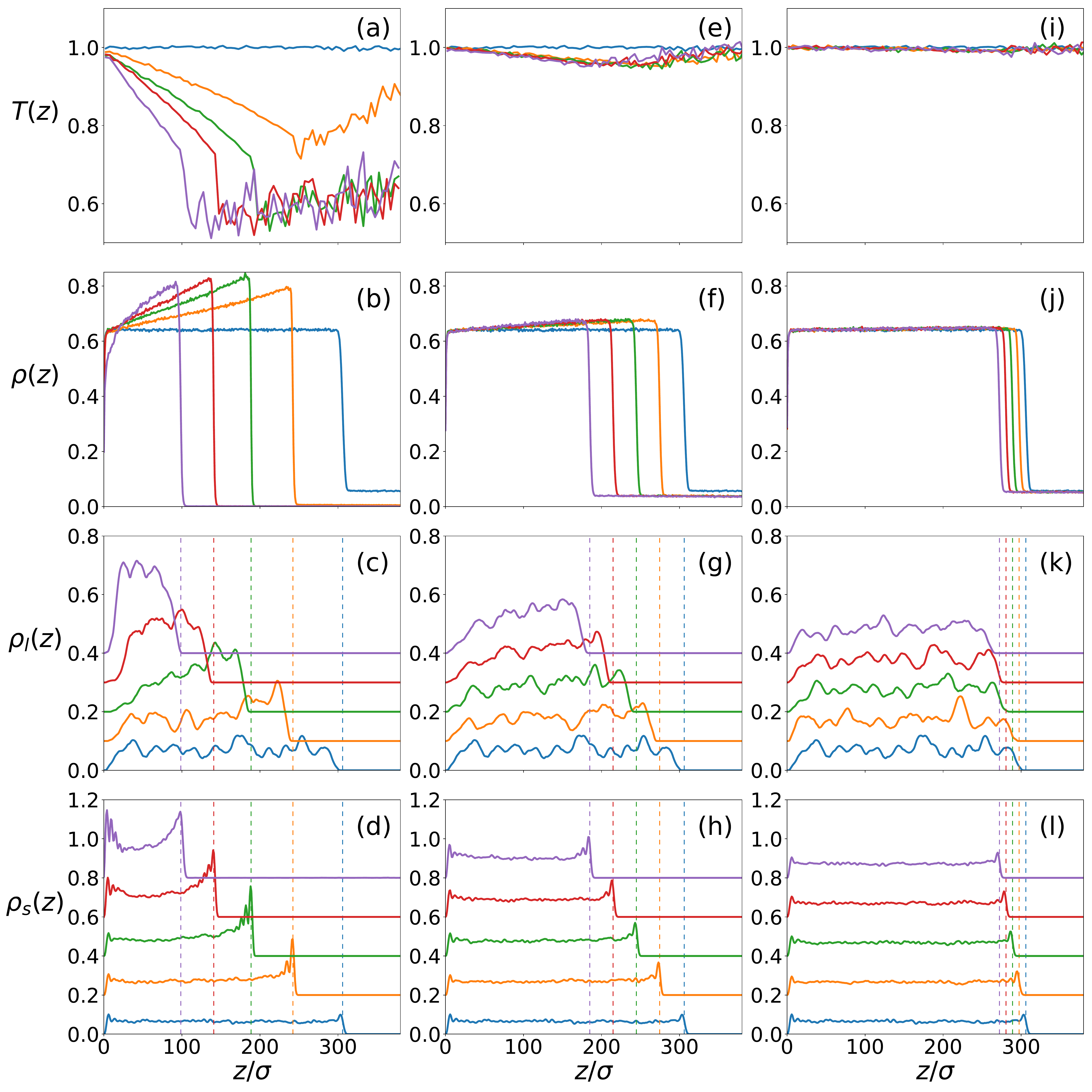}
\caption{Temperature and density profiles for $\phi_{0.068}R_{30}$ (left column), $\phi_{0.068}R_{5}$ (middle column), and $\phi_{0.068}R_{1}$ (right column): temperature (top row); solvent (second row); LNPs (third row); SNPs (bottom row). The vertical dashed lines indicate the location of the liquid/vapor interface. For clarity, the density profiles for LNPs (SNPs) are shifted upward successively by $0.1m/\sigma^{3}$ ($0.2m/\sigma^{3}$).}
\label{Figure:12800SNP_density}
\end{figure}

\begin{figure}[tp]
\includegraphics[width = 0.5\textwidth]{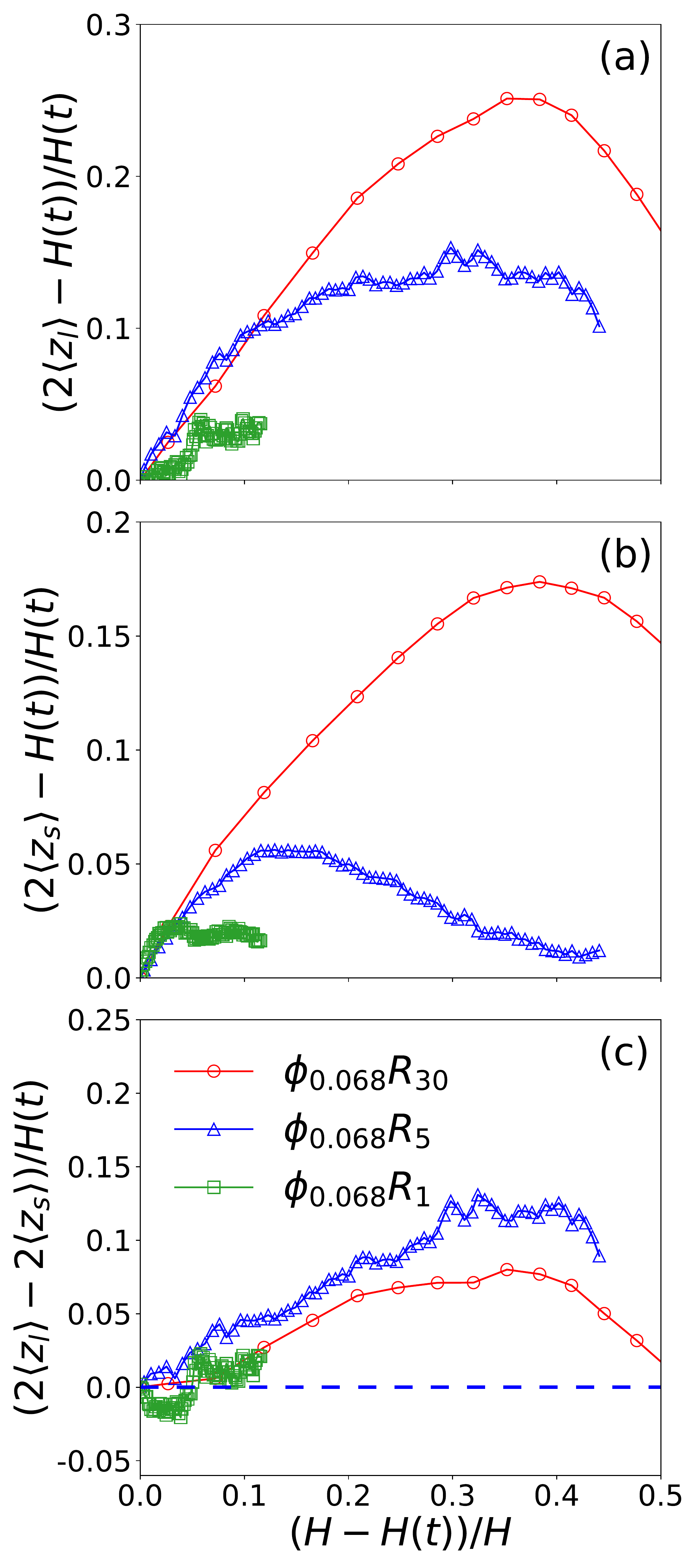}
\caption{Average position in the $z$ direction relative to the center of the film, normalized by $H(t)/2$, is plotted against time $(H-H(t))/H$ for (a) LNPs and (b) SNPs. Panel (c) shows the average separation between LNPs and SNPs, normalized by $H(t)/2$, as a function of $(H-H(t))/H$. Data are for system $\phi_{0.068}R_{30}$ (red circles), $\phi_{0.068}R_{5}$ (blue triangles) and $\phi_{0.068}R_{1}$ (green squares).}
\label{Figure:12800SNP_order}
\end{figure}

\noindent {\bf 12800-SNP Systems:} Figure~\ref{Figure:12800SNP_snapshots} shows snapshots of $\phi_{0.068}R_{30}$, $\phi_{0.068}R_{5}$, and $\phi_{0.068}R_{1}$ with $N_s=12800$ at various times during evaporation. The corresponding temperature and density profiles are shown in Fig.~\ref{Figure:12800SNP_density} and the average positions of NPs and their mean separation are shown in Fig.~\ref{Figure:12800SNP_order}. For these systems, $\phi_s=\phi_l=0.068$. The systems $\phi_{0.068}R_{30}$ and $\phi_{0.068}R_{5}$ behave similarly as the systems with $N_s = 1920$ and $6400$ and show the ``large-on-top'' stratification. However, the thickness of the interfacial region, in which the SNPs are accumulated and form a skin layer, becomes wider than those in the systems with $N_s = 1920$ and $6400$. Furthermore, the positive gradient of the density profile of LNPs below this skin layer of SNPs is smaller for $\phi_{0.068}R_{30}$ and $\phi_{0.068}R_{5}$ than for the systems with $N_s = 1920$ ($\phi_{0.011}R_{30}$ and $\phi_{0.011}R_{5}$) and $N_s = 6400$ ($\phi_{0.034}R_{30}$ and $\phi_{0.034}R_{5}$) at the same evaporation rate (i.e., the same $\zeta$ as in the subscript of $R$ in the system label). For $\phi_{0.068}R_{1}$, the density profiles of LNPs and SNPs are almost flat, except for the slight density peak of SNPs at the interface which exists even in equilibrium. We classify this system as ``uniform''. This classification is corroborated by the order parameter shown in Fig.~\ref{Figure:12800SNP_order}(c) for $\phi_{0.068}R_{1}$, which has a small magnitude and oscillates around 0.

In general at the same evaporation rate, the surface accumulation of SNPs becomes more significant and the accumulation of LNPs below the skin layer of SNPs becomes less significant when $\phi_s$ is increased. Eventually, the system is driven into the ``small-on-top'' regime when $\phi_s > c/{\rm Pe}_s$, where $c\sim 1$ and ${\rm Pe}_s$ is the P\'{e}clet number of SNPs. This transition is observed for the systems with $N_s = 19200$ ($\phi_{0.10}R_{30}$, $\phi_{0.10}R_{5}$, and $\phi_{0.10}R_{1}$), which are discussed in detail in the main text.

\begin{figure}[tp]
\includegraphics[width = 0.75\textwidth]{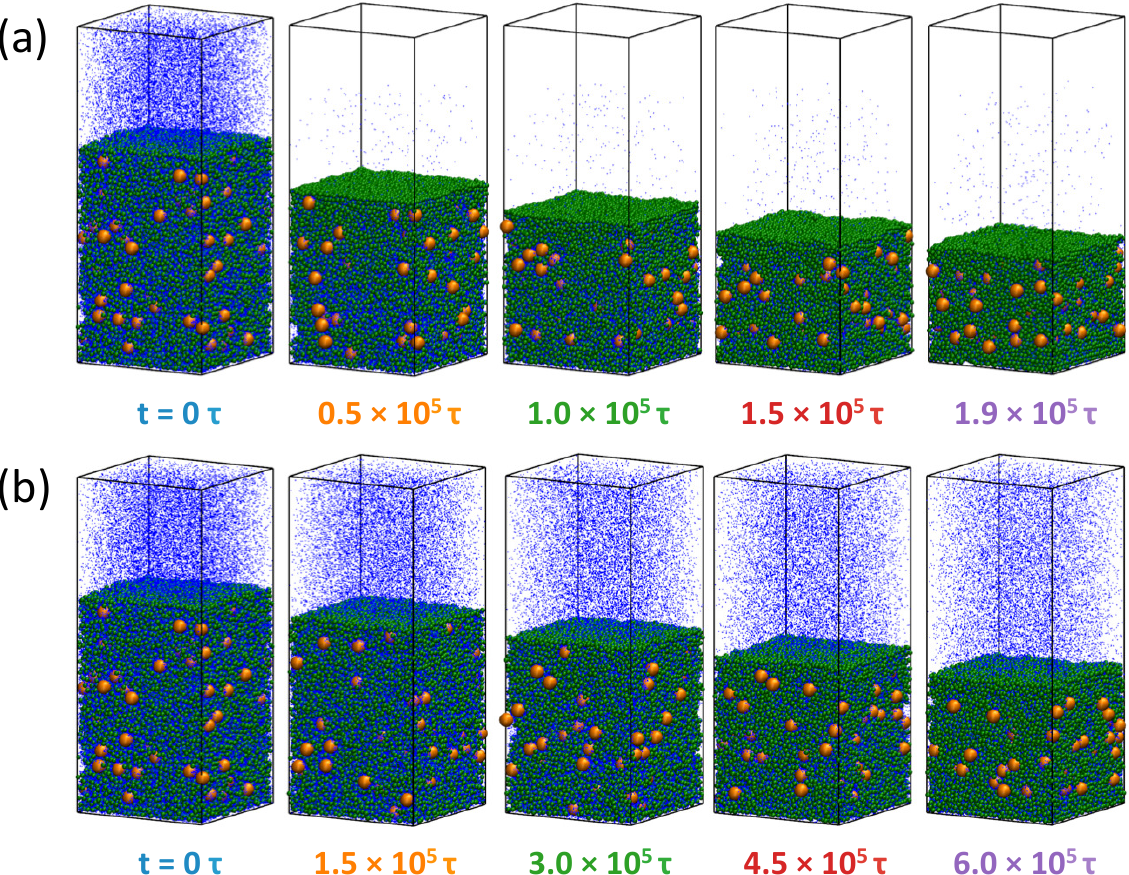}
\caption{Snapshots of (a) $\phi_{0.13}R_{30}$ and (b) $\phi_{0.13}R_{5}$ with $N_s = 25600$. Color code: LNPs (orange), SNPs (green), and solvent (blue). For clarity, only 5\% of the solvent beads are visualized. In the last frame the volume fractions of nanoparticles are: (a) $\phi_l$ = 0.13, $\phi_s$ = 0.26;(b) $\phi_l$ = 0.11, $\phi_s$ = 0.22.}
\label{Figure:25600SNP_snapshots}
\end{figure}

\begin{figure}[tp]
\includegraphics[width = 0.75\textwidth]{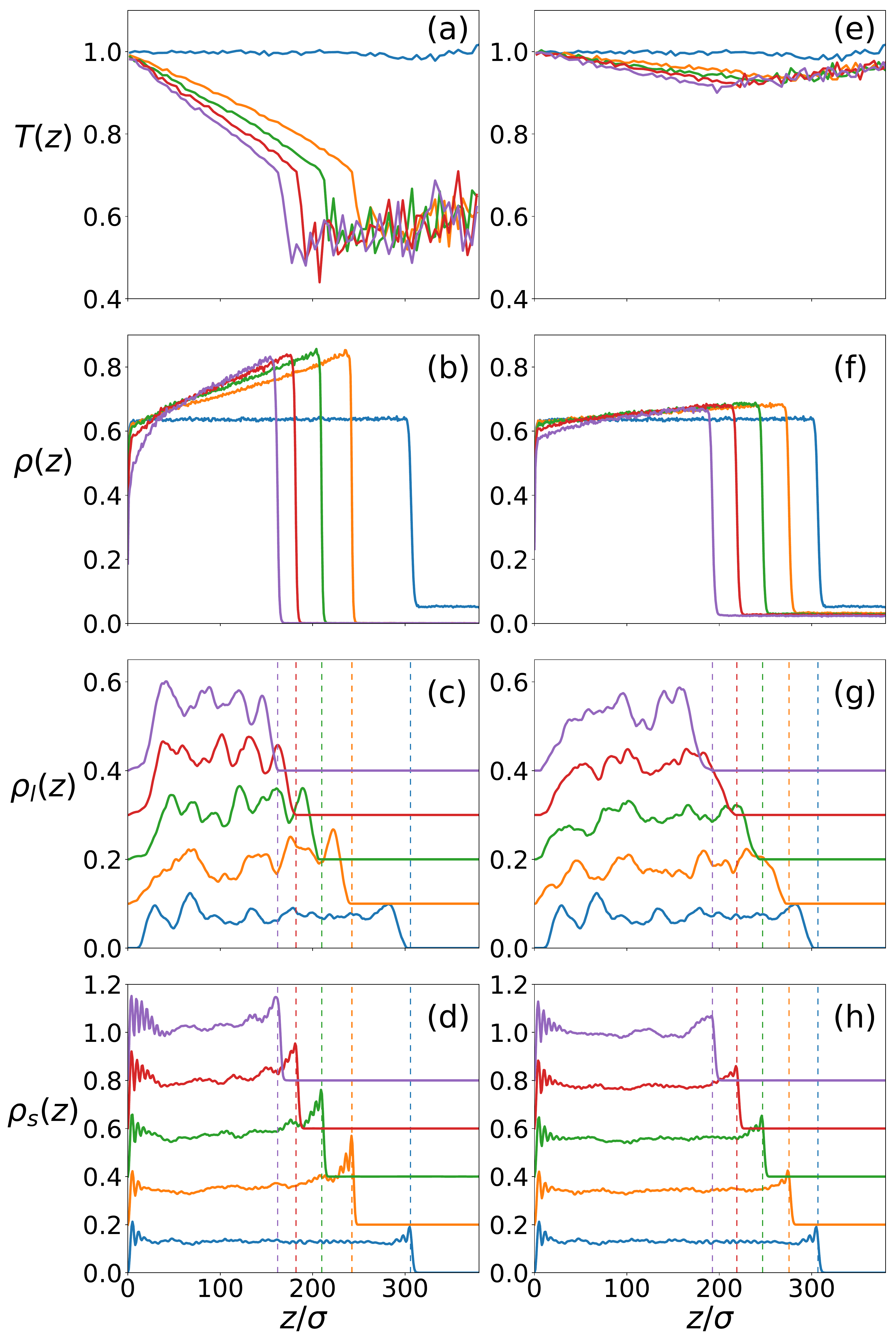}
\caption{Temperature and density profiles for $\phi_{0.13}R_{30}$ (left column) and $\phi_{0.13}R_{5}$ (right column): (a) and (e) temperature; (b) and (f) solvent; (c) and (g) LNPs; (d) and (h) SNPs. The vertical dashed lines indicate the location of the liquid/vapor interface. For clarity, the density profiles for LNPs (SNPs) are shifted upward successively by $0.1m/\sigma^{3}$ ($0.2m/\sigma^{3}$).}
\label{Figure:25600SNP_density}
\end{figure}

\begin{figure}[tp]
\includegraphics[width = 0.5\textwidth]{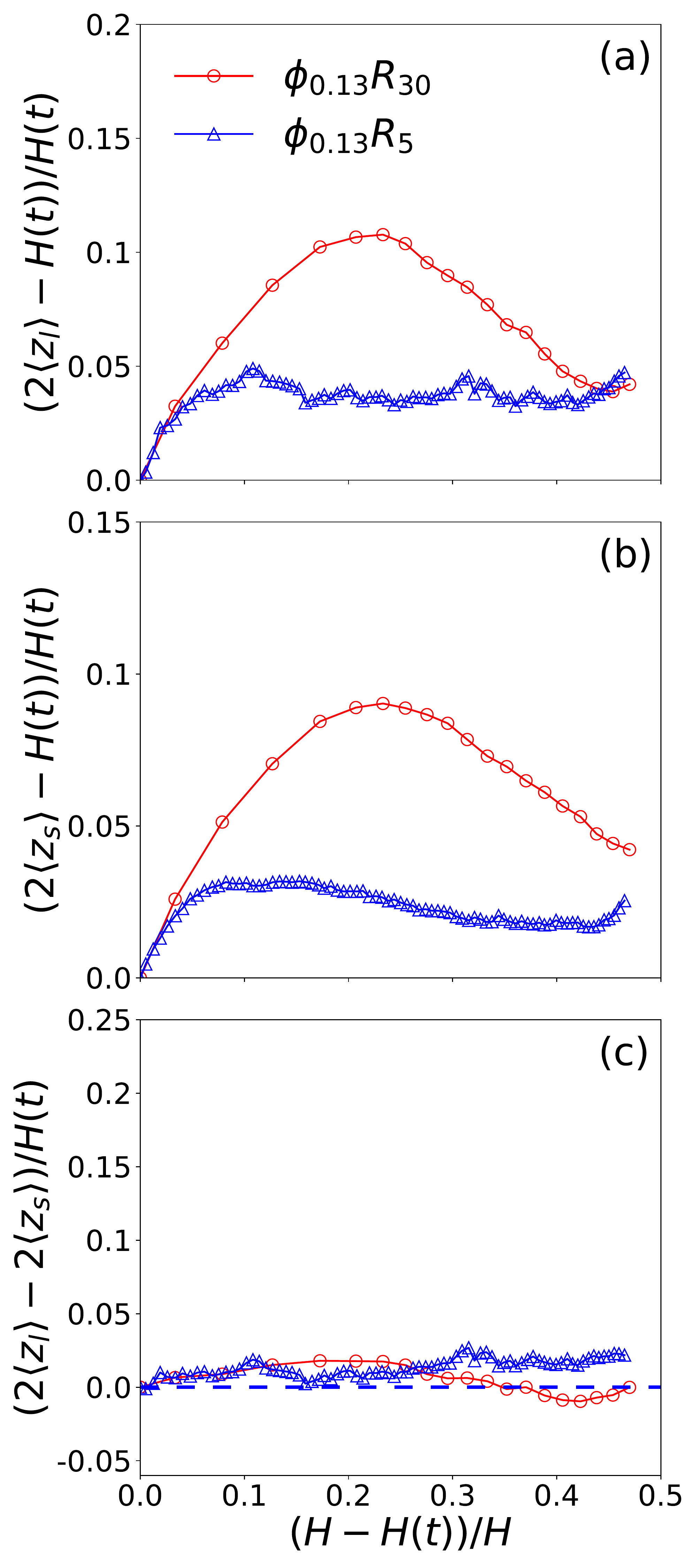}
\caption{Average position in the $z$ direction relative to the center of the film, normalized by $H(t)/2$, is plotted against time $(H-H(t))/H$ for (a) LNPs and (b) SNPs. Panel (c) shows the average separation between LNPs and SNPs, normalized by $H(t)/2$, as a function of $(H-H(t))/H$. Data are for system $\phi_{0.13}R_{30}$ (red circles) and $\phi_{0.13}R_{5}$ (blue triangles).}
\label{Figure:25600SNP_order}
\end{figure}

\noindent {\bf 25600-SNP and 32000-SNP Systems:} Figure~\ref{Figure:25600SNP_snapshots} shows snapshots of $\phi_{0.13}R_{30}$ and $\phi_{0.13}R_{5}$ with $N_s=25600$ at various times during evaporation. The corresponding temperature and density profiles are shown in Fig.~\ref{Figure:25600SNP_density} and the average positions of NPs and their mean separation are shown in Fig.~\ref{Figure:25600SNP_order}. For $\phi_{0.13}R_{30}$, the order parameter $ \langle z_l \rangle - \langle z_s\rangle $ first has a small positive value and then becomes negative. As shown in Figs.~\ref{Figure:25600SNP_density}(c), the density profiles of LNPs at early times have a small positive gradient, indicating that the LNPs accumulate slightly below the surface layer of SNPs. At late times, the density profiles of LNPs are almost flat. Fig.~\ref{Figure:25600SNP_density}(d) shows that when the interfacial region is approached, a positive gradient develops during evaporation for the density of SNPs. Combining the density profiles of LNPs and SNPs and the results on the order parameter, we classify $\phi_{0.13}R_{30}$ as ``small-on-top''. For $\phi_{0.13}R_{5}$, the density profiles of LNPs and SNPs during evaporation are almost always flat. We classify $\phi_{0.13}R_{5}$ as ``uniform'', though its order parameter has a small positive value.

\begin{figure}[tp]
\includegraphics[width = 0.75\textwidth]{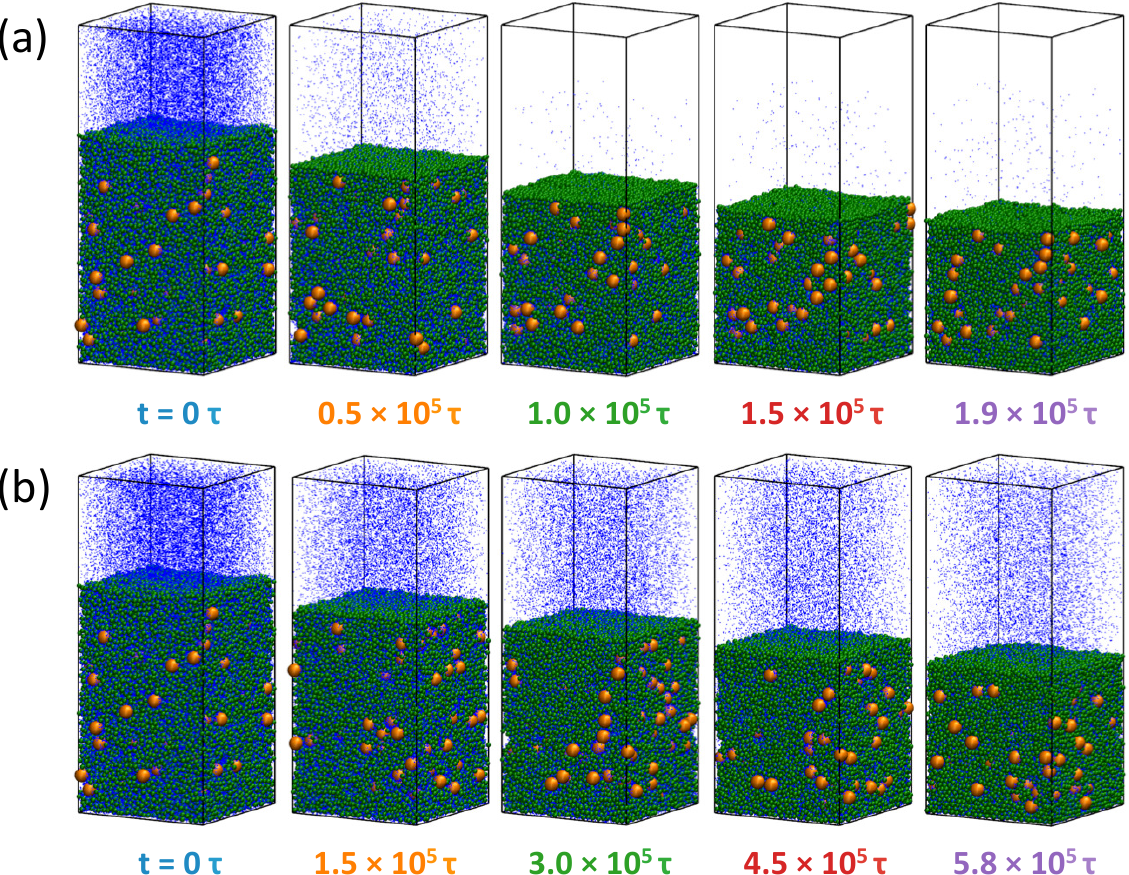}
\caption{Snapshots of (a) $\phi_{0.16}R_{30}$ and (b) $\phi_{0.16}R_{5}$ with $N_s = 32000$. Color code: LNPs (orange), SNPs (green), and solvent (blue). For clarity, only 5\% of the solvent beads are visualized. In the last frame the volume fractions of nanoparticles are: (a) $\phi_l$ = 0.11, $\phi_s$ = 0.26;(b) $\phi_l$ = 0.11, $\phi_s$ = 0.27.}
\label{Figure:32000SNP_snapshots}
\end{figure}

\begin{figure}[tp]
\includegraphics[width = 0.75\textwidth]{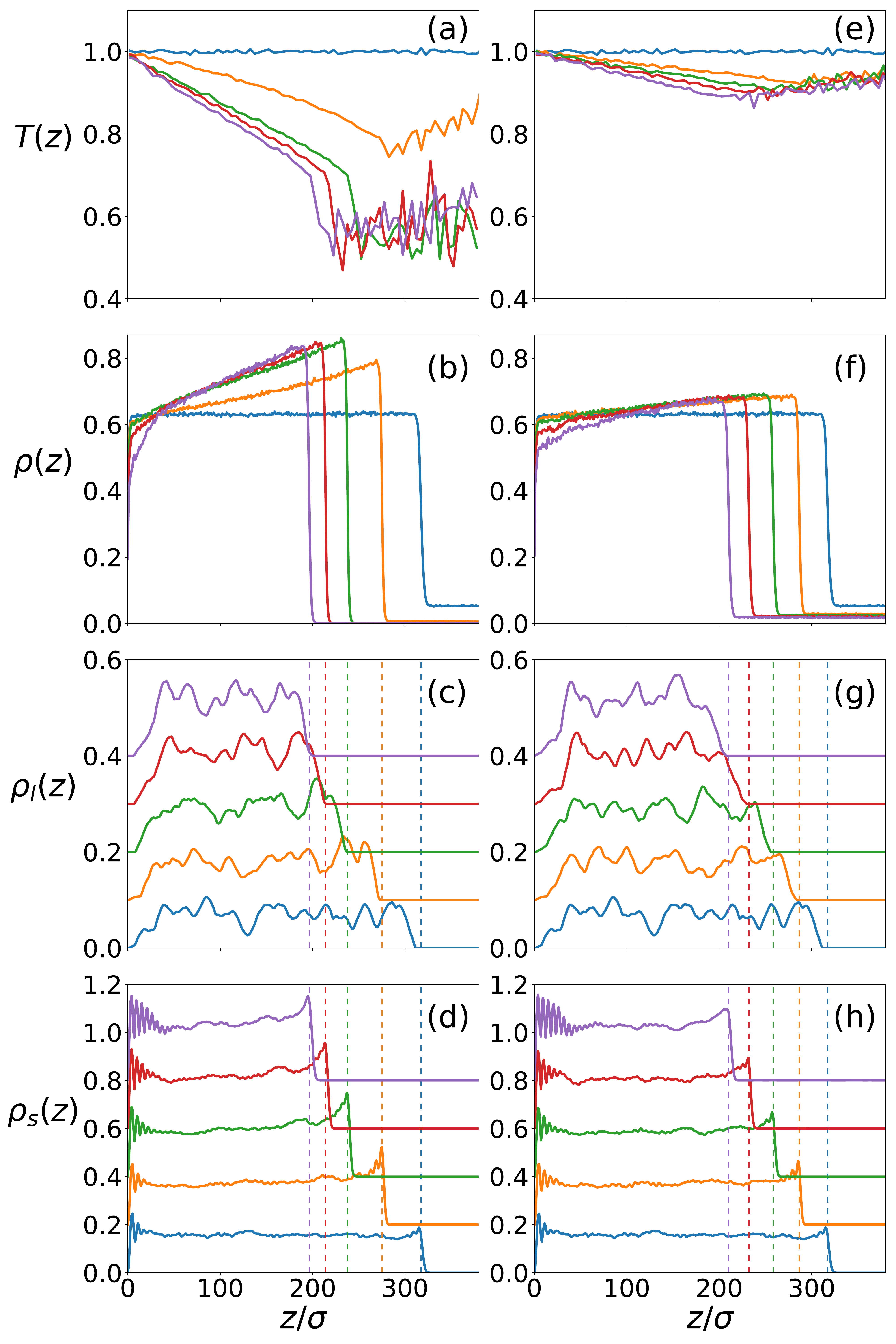}
\caption{Temperature and density profiles for $\phi_{0.16}R_{30}$ (left column) and $\phi_{0.16}R_{5}$ (right column): (a) and (e) temperature; (b) and (f) solvent; (c) and (g) LNPs; (d) and (h) SNPs. The vertical dashed lines indicate the location of the liquid/vapor interface. For clarity, the density profiles for LNPs (SNPs) are shifted upward successively by $0.1m/\sigma^{3}$ ($0.2m/\sigma^{3}$).}
\label{Figure:32000SNP_density}
\end{figure}

\begin{figure}[tp]
\includegraphics[width = 0.5\textwidth]{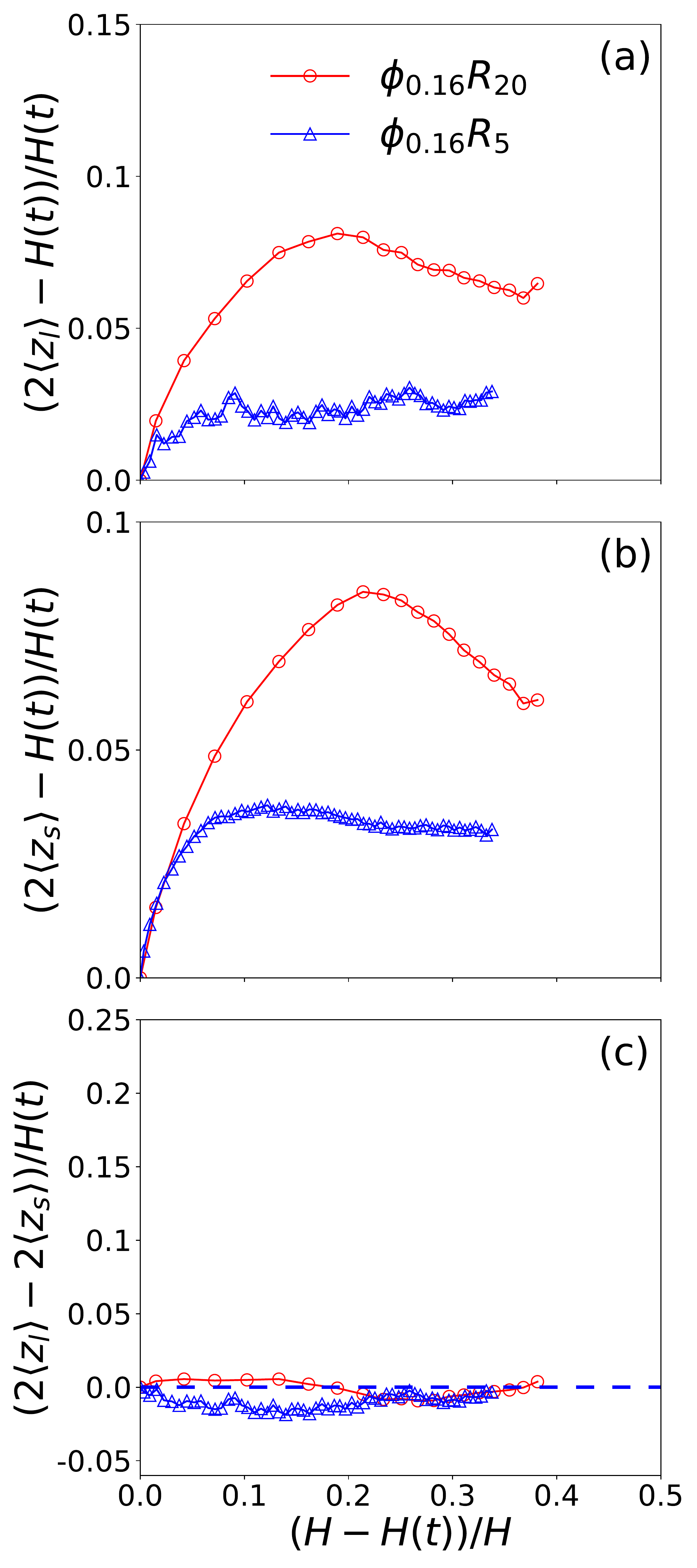}
\caption{Average position in the $z$ direction relative to the center of the film, normalized by $H(t)/2$, is plotted against time $(H-H(t))/H$ for (a) LNPs and (b) SNPs. Panel (c) shows the average separation between LNPs and SNPs, normalized by $H(t)/2$, as a function of $(H-H(t))/H$. Data are for system $\phi_{0.16}R_{30}$ (red circles) and $\phi_{0.16}R_{5}$ (blue triangles).}
\label{Figure:32000SNP_order}
\end{figure}

Figure~\ref{Figure:32000SNP_snapshots} shows snapshots of $\phi_{0.16}R_{30}$ and $\phi_{0.16}R_{5}$ with $N_s=32000$ at various times during evaporation. The corresponding temperature and density profiles are shown in Fig.~\ref{Figure:32000SNP_density} and the average positions of NPs and their mean separation are shown in Fig.~\ref{Figure:32000SNP_order}. These 2 systems behave similarly as the 2 systems with $N_s= 25600$. We classify $\phi_{0.16}R_{30}$ as ``small-on-top'' and $\phi_{0.16}R_{5}$ as ``uniform'', though the oder parameter in the latter system has a small negative value.

For the systems with $N_s=25600$ and $32000$, some ($\sim 500$) SNPs move into the vapor during solvent evaporation. This undesirable behavior prevented us from running systems with more SNPs. Exploring the regime with high volume fractions of SNPs remains an interesting direction for the future.

\noindent {\bf Evidence of Thermophoresis:} In the main text we argue that a positive gradient of the solvent density, which is induced by the evaporative cooling of the interface, drives NPs toward the interfacial region. This is an example of thermophoresis, where particle drift under a density gradient of the solvent caused by a temperature gradient.\cite{Brenner2011PRE} Our simulations of drying NP suspensions indicate that LNPs have a larger thermophoretic response to a gradient of the solvent density than SNPs. For example, the response of LNPs to the density profile of the solvent can be clearly seen in $\phi_{0.011}R_{30}$, where evaporative cooling is very strong and induces a gradient of the solvent density as shown in Fig.~\ref{Figure:1920SNP_density}(b). The migration of LNPs to the interface is distinct at $t=1\times10^5 \tau$ and $1.5\times 10^5 \tau$, as shown in Fig.~\ref{Figure:1920SNP_density}(c) as well as in Fig.~\ref{Figure:1920SNP_snapshots}(a). The similar trend, though weaker, persists to a lower evaporation rate as in $\phi_{0.011}R_{5}$ and higher volume fractions of SNPs as in $\phi_{0.034}R_{30}$ and $\phi_{0.034}R_{5}$.

\begin{figure}[tp]
\includegraphics[width = 0.5\textwidth]{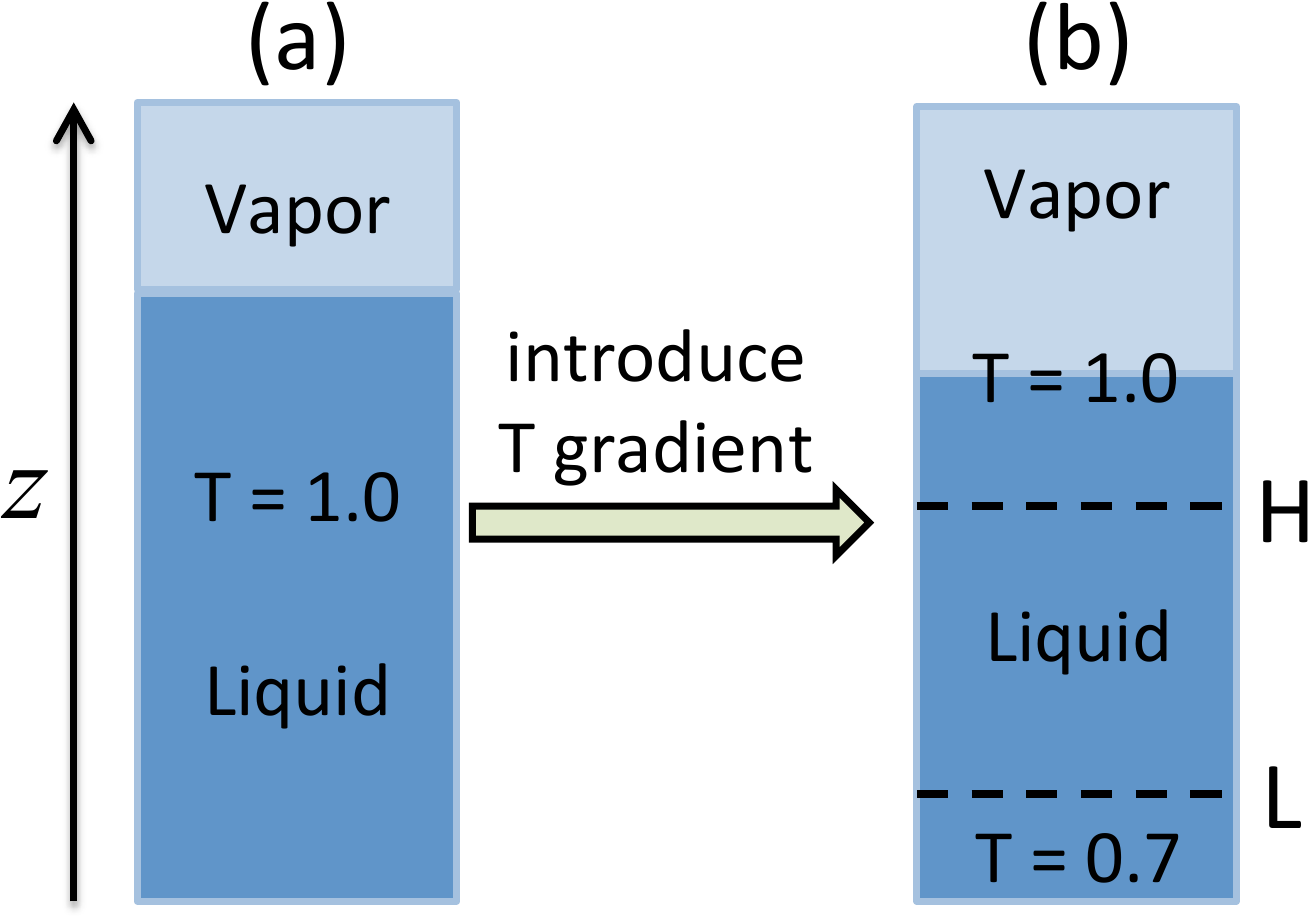}
\caption{The set-up of the simulation of thermophoresis: (a) the whole system is equilibrated at $T=1.0\epsilon/k_{\rm B}$; (b) a temperature gradient is introduced into the system by thermalizing the bottom of the liquid at $T=0.7\epsilon/k_{\rm B}$ while keeping the top of the liquid including the gas at $T=1.0\epsilon/k_{\rm B}$.}
\label{Figure:thermophoresis}
\end{figure}

\begin{figure}[tp]
\includegraphics[width = 0.5\textwidth]{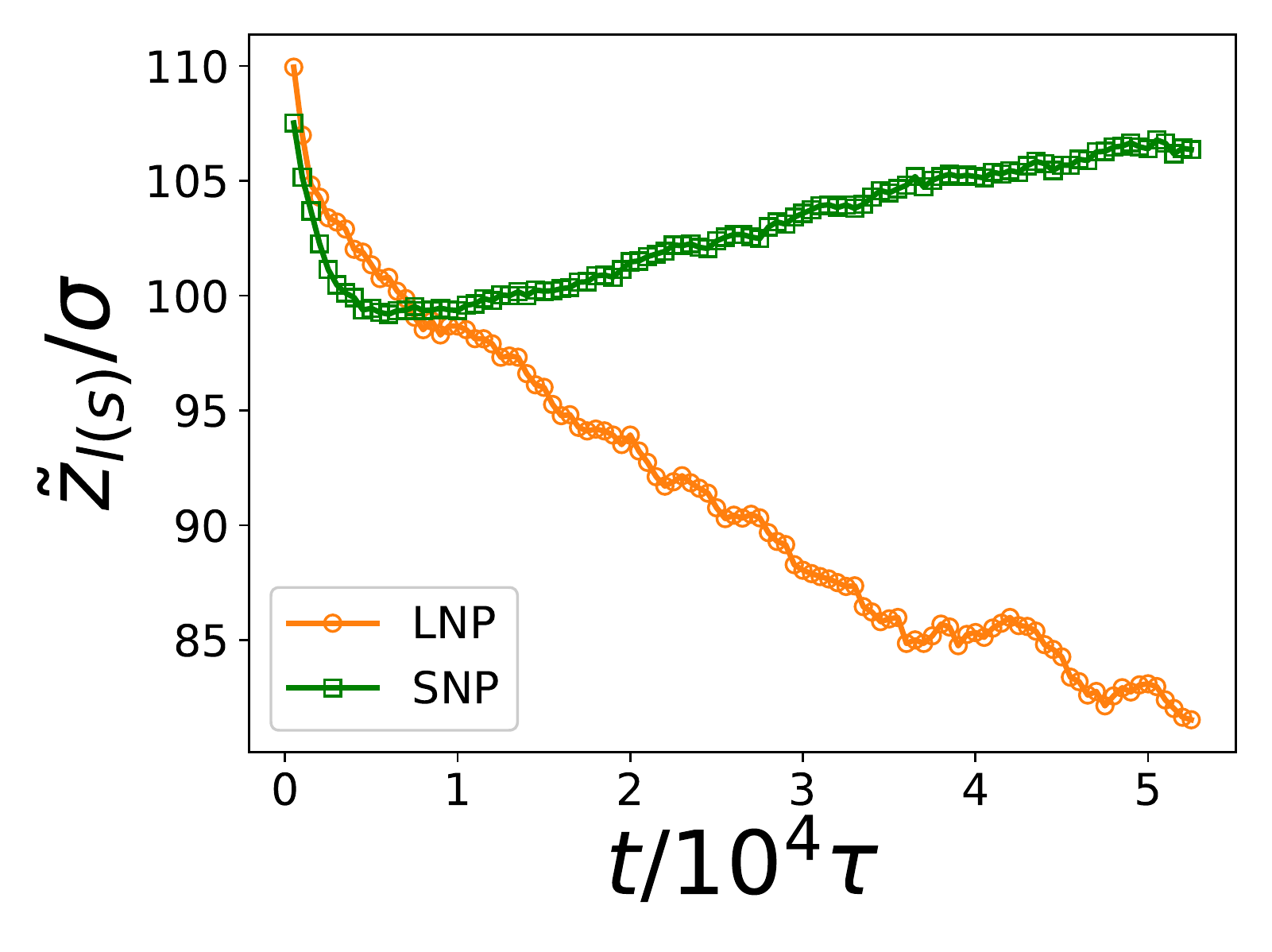}
\caption{The drift of LNPs and SNPs in response to a density gradient of the solvent induced by the temperature gradient introduced in Fig.~\ref{Figure:thermophoresis}(b).}
\label{Figure:NP_drift}
\end{figure}

We have performed an additional simulation to directly confirm thermophoresis, as illustrated in Fig.~\ref{Figure:thermophoresis}. We start with an equilibrium suspension of LNPs and SNPs with the whole system thermalized at $T=1.0\epsilon/k_{\rm B}$. In this system, the LNPs and SNPs are uniformly dispersed in the liquid solvent. Then a temperature gradient is introduced into the system by thermalizing a layer of liquid at the bottom of the suspension at $T=0.7\epsilon/k_{\rm B}$, while keeping the top of the liquid solvent including the gas at $T=1.0\epsilon/k_{\rm B}$. As a result, a positive temperature gradient is established along the $z$-direction, which induces a negative gradient of the solvent density. In Fig.~\ref{Figure:NP_drift}, the average locations of LNPs and SNPs along the $z$-direction, $\tilde{z}_l$ and $\tilde{z}_s$, are plotted against time $t$. Right after the temperature gradient is introduced into the system, the average density of the liquid decreases, which causes the liquid/vapor interface to recede. This is the reason that $\tilde{z}_l$ and $\tilde{z}_s$ first both decrease with $t$ quickly at early times. After this transient regime, the LNPs keep migrating toward the bottom of the suspension where the solvent density (temperature) is higher (lower), while the SNPs on average move toward the top of the liquid to some extent, which may be due to the diffusiophoretic force from the gradient of LNP concentrations or the back-flow in the solvent induced by LNP migration toward the bottom of the liquid. This contrasting behavior demonstrates that the thermophoretic effect is stronger for LNPs, which drift toward regions where the solvent density is higher. For a fast evaporating NP suspension, the temperature at the liquid/vapor interface is lower than that in the bulk and the solvent density there is higher. This density gradient pushes more LNPs toward the interfacial region, as discussed in the main text.

\begin{figure}[tp]
\includegraphics[width = 0.5\textwidth]{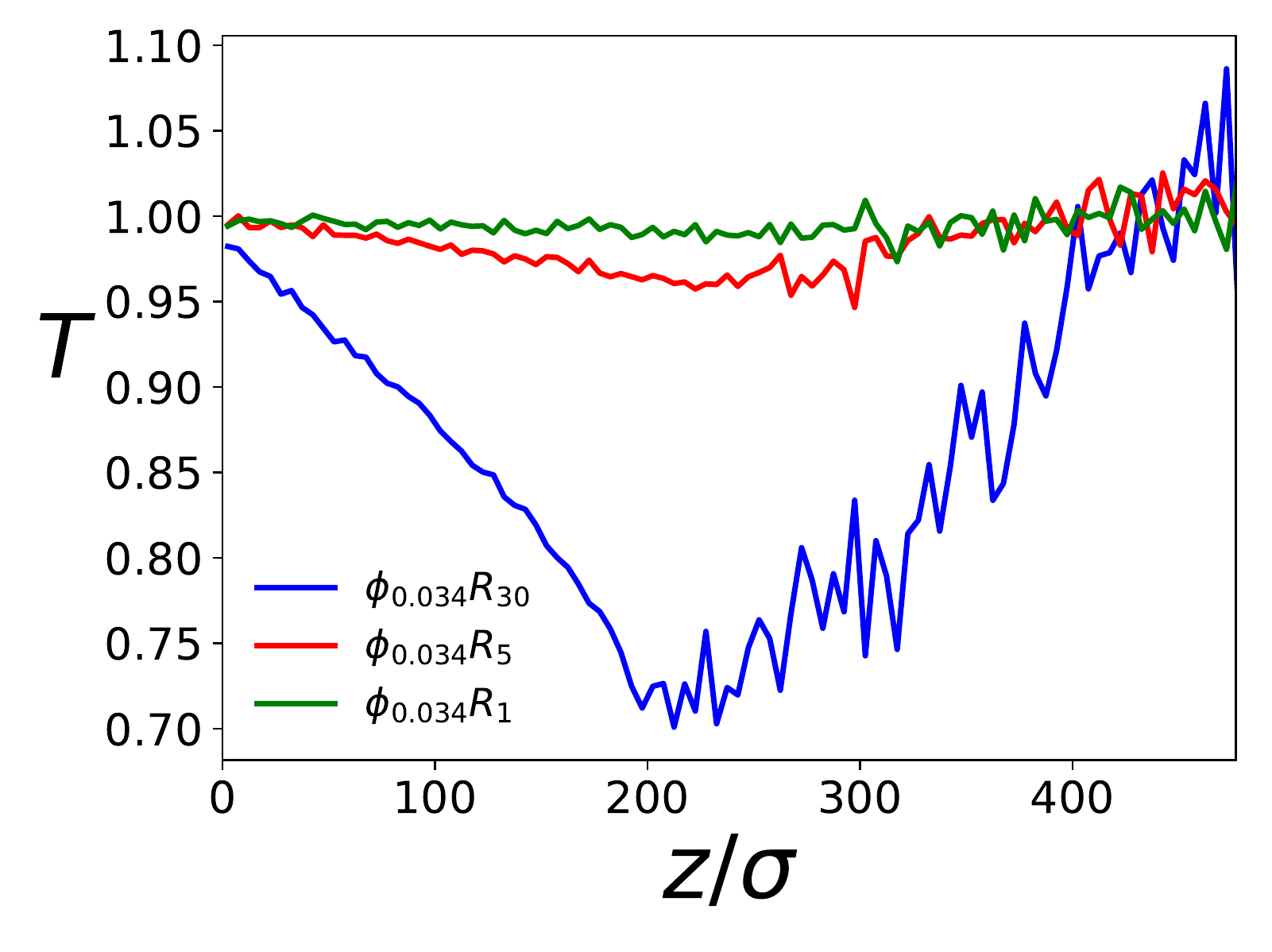}
\caption{The temperature profile along the $z$-direction at $t=1\times 10^5\tau$ for $\phi_{0.034}R_{30}$ (bottom blue line), $t=3\times 10^5\tau$ for $\phi_{0.034}R_{5}$ (middle red line), and $t=4\times 10^5\tau$ for $\phi_{0.034}R_{1}$ (top green line), respectively.}
\label{Figure:T_profile}
\end{figure}

\noindent {\bf Temperature and Pressure Profiles during Evaporation:} The gradient of the solvent density is induced by a temperature gradient in the solvent, which is caused by evaporative cooling at the interface as shown in the top rows of Figs.~\ref{Figure:1920SNP_density}, \ref{Figure:6400SNP_density}, \ref{Figure:12800SNP_density}, \ref{Figure:25600SNP_density}, and \ref{Figure:32000SNP_density}. We also show temperature profiles in Fig.~\ref{Figure:T_profile} for $\phi_{0.034}R_{30}$ at $t=1\times 10^5\tau$, for $\phi_{0.034}R_{5}$ at $t=3\times 10^5\tau$, and for $\phi_{0.034}R_{1}$ at $t=4\times 10^5\tau$, respectively, to illustrate their dependence on evaporation rates. The trend is clear: the higher the evaporation rate, the stronger the evaporative cooling and the associated temperature gradient in the liquid solvent.

\begin{figure}[tp]
\includegraphics[width = 0.5\textwidth]{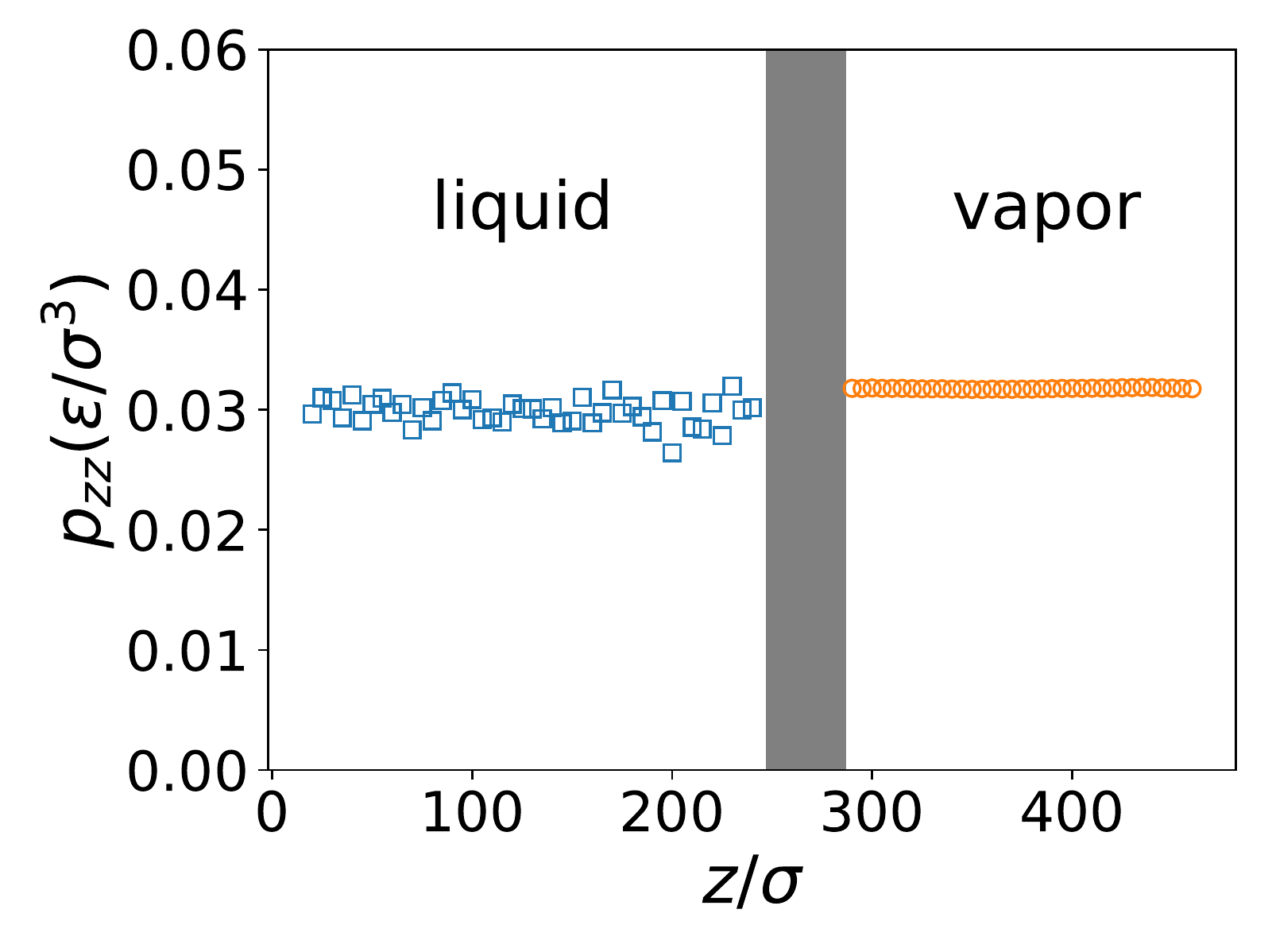}
\caption{$P_{zz}$ along the $z$-direction at $t=2\times 10^5\tau$ for $\phi_{0.034}R_{5}$.}
\label{Figure:Pzz_profile}
\end{figure}

There is no pressure gradient in the evaporating suspensions studied here as mechanical equilibrium is established very quickly even for ultrafast evaporation.\cite{Cheng2011} One example of the $zz$-component of the pressure tensor, $p_{zz}$, along the $z$-direction is shown in Fig.~\ref{Figure:Pzz_profile} for $\phi_{0.034}R_{5}$ at $t=2\times 10^5\tau$. The pressure tensor is computed using the Irving-Kirkwood formula,\cite{Irving1950}
\begin{equation}
p_{\alpha\beta} = \frac{1}{V} \Bigg( 
\sum_{j} m_j v_{j \alpha} v_{j \beta} - \frac{1}{2}\sum_{i \ne j} \sum_j r_{ij\alpha} \frac{\partial \varphi_{ij}}{\partial r_{ij\beta}}
\Bigg),
\end{equation}
where $j$ indexes all particles inside the volume $V$, the summation over $i$ is over all particles in the system (i.e., the second summation is over all particle pairs with at least one particle inside $V$),  $\varphi_{ij}$ is the pairwise potential between particles $i$ and $j$ as a function of their separation $r_{ij}$, $m_j$ and $v_j$ are the mass and velocity of the $j$-th particle, and Greek indices ($\alpha$ and $\beta$) indicate components in the $x$, $y$, and $z$ directions, respectively. The results clearly show that the pressure is constant in the direction normal to the liquid-vapor interface. Detailed discussion of temperature and pressure profiles in an evaporating liquid can be found in Ref.~\onlinecite{Cheng2011}.

\noindent {\bf Movies of Evolution of Density Profiles during Evaporation:} The evolution of density profiles of the solvent, LNPs, and SNPs for each system is uploaded as movie files ``Movie\_XXXSNP--$\zeta$.mp4'', where ``XXX'' indicates the number of SNPs and $\zeta$ is the number of vapor particles removed in the deletion zone every $\tau$.

\bibliography{../evap}

\begin{thebibliography}{44}
\bibitem[{ACS({\natexlab{a}})}]{ACSNano2018Note1}
\bibinfo{note}{The simulation results with $\zeta = 30$ or 20 are very close to
  those obtained with the solvent evaporating into a vacuum.}
\bibitem[{ACS({\natexlab{b}})}]{ACSNano2018Note2}
\bibinfo{note}{The radius of the spherical excluded volume of a SNP (LNP) is
  determined to be $3.12\sigma$ ($10.6\sigma$), which is $0.62\sigma$
  ($0.6\sigma$) larger than its nominal radius.}
\bibitem[{ACS({\natexlab{c}})}]{ACSNano2018Note3}
\bibinfo{note}{We have observed that some SNPs ($\sim 500$) moved into the
  vapor during solvent evaporation for $\phi_s = 0.13$ and 0.16. These SNPs are
  exlcuded from the analyses presented here.}
\end{thebibliography}

\newpage

\end{document}